\documentclass[letterpaper,12pt]{article}
\pdfoutput=1 

\usepackage{jheppub} 
\usepackage{amsmath,amsfonts,amssymb}
\usepackage{psfrag}
\usepackage{enumerate}
\usepackage{mathrsfs}
\usepackage{graphicx}
\usepackage{wrapfig}
\usepackage{xcolor}
\usepackage{caption}

\numberwithin{equation}{section}

\makeatletter
\def\@fpheader{\relax}
\makeatother

\newcommand{\be}{\begin{equation}}
\newcommand{\ee}{\end{equation}}
\newcommand{\beq}{\begin{eqnarray}}
\newcommand{\eeq}{\end{eqnarray}}

\def\[{\left [}
\def\]{\right ]}
\def\({\left (}
\def\){\right )}

\def\r2{\sqrt{2}}

\newcommand{\bbibitem}[1]{\bibitem{#1}\marginpar{#1}}

\newcommand{\myeq}[1]{\begin{equation} #1 \end{equation}}
\newcommand{\myal}[1]{\begin{align} #1 \end{align}}
\newcommand{\OO}{\mathcal{O}}

\def\Label#1{\label{#1}%
  \smash{\hbox to0pt{\raise1ex\hbox{\tiny[#1]}\hss}}}
\def\noLabels{\let\Label=\label}
\def\nobbibitem{\let\bbibitem=\bibitem}

\subheader{SU-ITP-16/07}
\title{A Stereoscopic Look into the Bulk}

\author[a]{Bart{\l}omiej Czech,}
\author[a]{Lampros Lamprou,}
\author[a]{Samuel McCandlish,}
\author[a]{Benjamin Mosk,}
\author[b]{and James Sully}

\affiliation[a]{Stanford Institute for Theoretical Physics, Department of Physics, Stanford University\\
Stanford, CA 94305, USA}
\affiliation[b]{Theory Group, SLAC National Accelerator Laboratory\\ Menlo Park,
CA 94025, USA}

\emailAdd{czech@stanford.edu}
\emailAdd{llamprou@stanford.edu}
\emailAdd{smccandlish@stanford.edu}
\emailAdd{bmosk1@stanford.edu}
\emailAdd{jsully@slac.stanford.edu}

\abstract{We present the foundation for a holographic dictionary with depth perception.  The dictionary consists of natural CFT operators whose duals are simple, diffeomorphism-invariant bulk operators. The CFT operators of interest are the ``OPE blocks,'' contributions to the OPE from a single conformal family. In holographic theories, we show that the OPE blocks are dual at leading order in $1/N$ to integrals of effective bulk fields along geodesics or homogeneous minimal surfaces in anti-de Sitter space.  One widely studied example of an OPE block is the modular Hamiltonian, which is dual to the fluctuation in the area of a minimal surface. Thus, our operators pave the way for generalizing the Ryu-Takayanagi relation to other bulk fields.

Although the OPE blocks are non-local operators in the CFT, they admit a simple geometric description as fields in kinematic space---the space of pairs of CFT points.  We develop the tools for constructing local bulk operators in terms of these non-local objects. The OPE blocks also allow for conceptually clean and technically simple derivations of many results known in the literature, including linearized Einstein's equations and the relation between conformal blocks and geodesic Witten diagrams.}

\begin{document}
\noLabels 
\nobbibitem 

\maketitle
\flushbottom
\tableofcontents

\section{Introduction}

This paper proposes a natural operator basis for conformal field theories, one that is particularly keen-sighted when used to view bulk physics in the AdS/CFT correspondence \cite{Maldacena:1997re}. We show these operators are both a powerful tool for performing calculations in AdS/CFT, and also suggestive of the right organizational structure for understanding gravitational physics. 

To begin, let us ask: what properties would characterize a natural set of holographic CFT variables and their duals?
\begin{itemize}
\item In the bulk, we should demand {\bf diffeomorphism invariance}. This seems to eliminate local quantities in favor of extended objects that reach out to the asymptotic boundary \cite{Maldacena:1998im,Susskind:1999ey,Horowitz:2009wm}.
\item On the CFT side, we should require a {\bf nice transformation law under conformal symmetry}. This will also ensure a corresponding covariance under AdS isometries.
\item Our variables should have an aesthetic appeal on both sides, 
even without reference to holography.
\end{itemize}

A prototypical example of such natural variables is encapsulated by the Ryu-Takayanagi proposal \cite{Ryu:2006bv,Ryu:2006ef}. In AdS, minimal surfaces are simple, diffeomorphism invariant, extended objects which reach out to the asymptotic boundary. On the CFT side, they find a compelling interpretation in terms of entanglement entropies, whose UV divergences transform covariantly under conformal symmetries. These properties allowed the RT proposal to revolutionize our understanding of holographic duality. 

The present paper takes seriously the lesson from Ryu-Takayanagi and organizes the AdS/CFT operator dictionary according to similar guidelines. 
In the CFT, we propose the right quantity is an ``OPE block,'' a well-known class of simple, but non-local, operators that are singled out by conformal symmetry. We show that these OPE blocks are holographically dual to bulk operators smeared along geodesics. This duality can be understood as an operator generalization of the Ryu-Takayanagi proposal.

To reach this conclusion, the first step is the observation that scale must be a key ingredient on the CFT side; without it, we will never probe the bulk. This automatically disqualifies local operators in the CFT. The simplest way to proceed is to consider CFT bi-locals, pairs of operator insertions whose separation coordinatizes the scale direction. Observing the bulk from two boundary viewpoints at a time will give us the benefit of stereoscopic vision: it will provide a sense of depth.\footnote{Some readers may be quick to interject (correctly) that CFT bi-locals do not seem sufficiently non-local to probe the infrared bulk geometry in any meaningful sense. Such an astute reader is asked to be patient.} Indeed, pairs of CFT points select natural, diffeomorphism invariant, extended bulk objects---geodesics.

We are next led to ask: how should we organize CFT bi-locals? The obvious answer is the operator product expansion (OPE). The kinematics of conformal invariance picks out a preferred basis of operators for the OPE, which we call ``OPE blocks.''\footnote{Now patience's reward: unlike the bi-local operator itself, the OPE block is not well-localized at any one (or two) boundary points.} 
While these are already well-known objects in the study of CFT, 
we will show that they also appear naturally in the study of entanglement. 
The modular Hamiltonian \cite{Casini2011} is precisely an OPE block. 
Most importantly for our argument, we show that the OPE block conformal kinematics can be equivalently written as a Klein-Gordon equation in the space of CFT bi-locals, what we call \emph{kinematic space} \cite{Czech:2014ppa, Czech:2015qta, Czech:2015kbp} (see \cite{PhysRevLett.116.061602} for the same observation restricted to the modular Hamiltonian and higher-spin charges). 

The appearance of kinematic space facilitates our derivation of the bulk dual of an OPE block: the kinematic space of CFT bi-locals is simultaneously the space of bulk geodesics. We will show that operators smeared over bulk geodesics obey the same equations of motion, constraints, and boundary conditions in kinematic space as do the OPE blocks. OPE blocks and geodesic operators can thus be understood as the same local kinematic operators. 

In AdS$_{>3}$ the bulk story becomes even richer because two time-like separated boundary points select a homogeneous codimension-2 surface rather than a geodesic. In an effort to minimize distractions, we will postpone a discussion of the higher-dimensional story until Sec.~\ref{higherd} and focus in most of the paper on AdS$_3$, where our results are easiest to state.

Our formalism unifies and contextualizes many important results, which were previously reported in the literature under diverse contexts. This includes:
\begin{itemize}
  \setlength{\itemsep}{2pt}
  \setlength{\parskip}{0pt}
	\item a new construction for local bulk operators \cite{Banks:1998dd,Balasubramanian:1998de,Bena:1999jv,Hamilton:2005ju,Hamilton:2006az,Heemskerk:2012np,Heemskerk:2012mn,Nakayama:2015mva};
	\item a novel look at the modular Hamiltonian \cite{Casini2011}; 
	\item the origin of geodesic Witten diagrams and their use in computing conformal blocks \cite{Hijano:2015,Hijano:2015zsa,HijanoKrausPerlmutterEtAl2015}; 
	\item and a re-derivation of Einstein's equations from entanglement (along the lines of \cite{Lashkari:2013koa,Faulkner:2013ica,Swingle:2014uza}),  whose details we largely leave to a forthcoming publication \cite{Czech:ee}. 
\end{itemize}
We devote Sec.~\ref{localbulkops} to local bulk operators and Sec.~\ref{apps} to the remaining applications.

In summary, our paper introduces a new entry to the holographic dictionary. 
On the CFT side, in Sec.~\ref{opeblocks} we define ``OPE blocks,'' a natural operator basis suggested by the operator product expansion. 
We explain in Sec.~\ref{gops} that their holographic duals are bulk operators integrated along geodesics. After discussing the applications of our dictionary, the paper closes with a summary of the story in higher dimensions (Sec.~\ref{higherd}), a Discussion section and three appendices, where we collect useful technicalities. 
\\

During this project we learned that another group---de Boer, Haehl, Heller and Myers---have been working on an overlapping set of ideas. Their paper on the subject will appear shortly \cite{MdBHH}.

\subsection{The Kinematic Space of \texorpdfstring{AdS$_3$}{AdS3}}
\label{ads3kin}

The stage on which our story unfolds is the kinematic space of AdS$_3$, which we presently discuss. This is more than a review of \cite{Czech:2015qta}, because that work was only concerned with geodesics living on a static slice of AdS$_3$. In this paper, where we make extensive use of conformal symmetry, restricting to a time slice would be unnecessarily limiting. 

We define kinematic space to be the space of ordered pairs of CFT points.  We will see, however, that kinematic space for $\rm{AdS}_3/\rm{CFT}_2$ can also be thought of as the space of any of the following objects:
\begin{itemize}
\item Causal diamonds $\diamond_{12}$ in the CFT
\item Pairs of time-like separated points that live on the remaining corners of $\diamond_{12}$
\item Oriented $\rm{AdS}_3$ geodesics $\gamma_{12}$, which asymptote to boundary points $x_1$ and $x_2$
\end{itemize}

\begin{figure}
\centering{\includegraphics[width=0.9\textwidth]{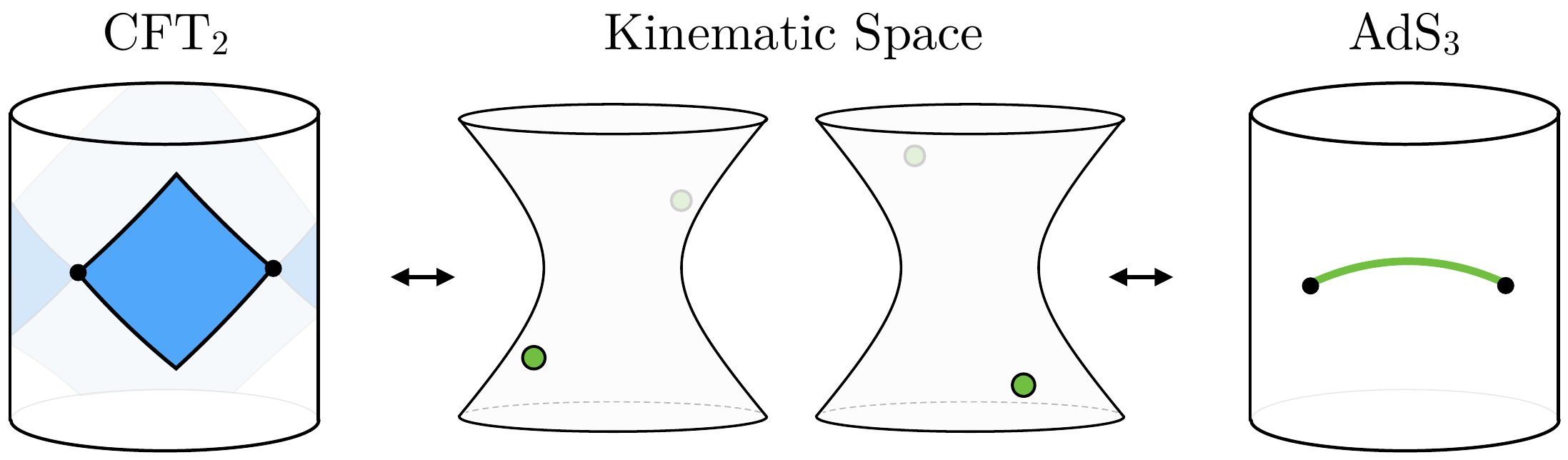}
\caption{Kinematic space for $\rm{AdS}_3$ has the $\rm{dS}_2 \times \rm{dS}_2$ metric of eq.~(\ref{eq:ksads3}). It is both the space of causal diamonds in $\rm{CFT}_2$ and the space of space-like geodesics in $\rm{AdS}_3$. To account for complementary causal diamonds (such as the two shown in the left panel) which are associated with the same geodesic (the right panel), we work with the space of oriented geodesics. The middle panel shows the two images of the same bulk geodesic that differ in orientation.}
\label{fig:ads3-ks-diagram}}
\end{figure}

Thinking of this kinematic space as comprising pairs of CFT points suggests natural coordinates on it: $x_1, x_2 \in {\rm CFT}$. In the AdS$_3$/CFT$_2$ correspondence, we are therefore looking at a four-dimensional space. When $x_1$ and $x_2$ are space-like separated, they are connected by a unique geodesic in AdS$_3$. 
This case ought to be distinguished from timelike separated $x_1$ and $x_2$, which are not endpoints of any bulk geodesic. The convenience of CFT$_2$ is that this distinction is immaterial: two time-like separated points instead define a causal diamond, whose spacelike separated corners again select a bulk geodesic. Thus, the kinematic space is really a space of boundary causal diamonds, each of which is canonically related to a unique spacelike geodesic in AdS$_3$ (see Fig.~\ref{fig:ads3-ks-diagram}).\footnote{Note that a geodesic maps not to one but to two complementary causal diamonds. For this reason, it is often convenient to define kinematic space as comprising \emph{oriented} geodesics, which do pick out a unique diamond.}  In higher dimensions, however, pairs of spacelike and timelike separated points give rise to genuinely distinct spaces and must be treated separately. Between now and Sec.~\ref{apps} we largely ignore this subtlety, postponing an account of higher-dimensional spaces to Sec.~\ref{higherd}.

\paragraph{Metric} Conformal symmetry picks out a unique metric for this kinematic space. To see this, consider the distance between two neighboring kinematic elements, $(x_1,x_2)$ and $(x_1+dx_1, x_2+dx_2)$:
\begin{equation}
ds^2=f_{\mu\nu}\left(x_1,x_2\right)\,dx_1^{\mu}dx_2^{\nu}\,,
\end{equation}
No cross-terms $dx_1^{\mu}dx_1^{\nu}$ or $dx_2^{\mu}dx_2^{\nu}$ appear because no invariant cross-ratio can be formed from the coordinates of three boundary points; we must move both $x_1$ and $x_2$ to obtain a nonzero distance. Now note that a conformal map $\left(x_1,x_2\right)\rightarrow\left(x_1^{\prime},x_2^{\prime}\right)$ transforms
$f_{\mu\nu}\left(x_1,x_2\right)$ as
\begin{equation}
f_{\mu\nu}\left(x_1,x_2\right)\rightarrow\frac{dx_1^{\prime \alpha}}{dx_1^{\mu}}\frac{dx_2^{\prime\beta}}{dx_2^{\nu}}f_{\alpha\beta}\left(x_1^{\prime},x_2^{\prime}\right).
\end{equation}
This is the transformation rule for a vacuum two-point function
of spin-1 CFT quasiprimaries with scaling dimension $\Delta=1$. The standard result for the two-point function is
\begin{equation}
\left\langle \mathcal{O}_{\mu}\left(x_1\right)\mathcal{O}_{\nu}\left(x_2\right)\right\rangle \propto\frac{I_{\mu\nu}\left(x_1-x_2\right)}{\left|x_1-x_2\right|^{2\Delta}},
\end{equation}
where the matrix $I_{\mu\nu}$ is fixed by symmetry to be \cite{Osborn:1993cr}:
\begin{equation}
I_{\mu\nu}(x)\equiv \eta_{\mu\nu}-2\frac{x_{\mu}x_{\nu}}{x^{2}}.
\label{invmatrix}
\end{equation}
In the end, the metric on kinematic space becomes:
\begin{equation}
ds^{2}=4\frac{I_{\mu\nu}\left(x_1-x_2\right)}{\left|x_1-x_2\right|^{2}}dx_1^{\mu}dx_2^{\nu}.\label{eq:kinmetric}
\end{equation}
where we have chosen the overall coefficient for later convenience.  Because this derivation does not use any facts specific to CFT$_2$, we will be able to re-use metric~(\ref{eq:kinmetric}) in Sec.~\ref{higherd} where we discuss kinematic spaces of higher-dimensional anti-de Sitter geometries.

\paragraph{Coordinates and factorization}
Note that the coordinates $x_1^\mu$ and $x_2^\nu$ in metric~(\ref{eq:kinmetric}) form two light-like pairs, so the signature of the AdS$_3$ kinematic space is $(2,2)$.\footnote{It will be $(d,d)$ for AdS$_{d+1}$; see Sec.~\ref{higherd}.} 
This fact is independent of how we choose $x_1^\mu$ and $x_2^\nu$ within the CFT: the spatial coordinate of $x_1$ matches up with the spatial coordinate of $x_2$ to form one pair of light-like coordinates on the kinematic space while the temporal coordinates of $x_1$ and $x_2$ form the other light-like pair. Yet one chart for $x_1$ and $x_2$ is more convenient than others: the coordinates $z_1 = t_1+x_1, \bar{z}_1 = t_1 - x_1$ (respectively $z_2, \bar{z}_2$) that are light-like \emph{in the CFT}. 

Substituting these in (\ref{eq:kinmetric}) gives:
\begin{equation}
ds^{2}=\frac{1}{2}\left[\frac{dz_1 dz_2}{\left(\frac{z_1-z_2}{2}\right)^{2}}+\frac{d\bar{z}_1d\bar{z}_2}{\left(\frac{\bar{z}_1-\bar{z}_2}{2}\right)^{2}}\right] = \frac{1}{2}\left[ds_{z}^{2}+ds_{\bar{z}}^{2}\right].
\label{eq:ksads3}
\end{equation}
We find a sum of two two-dimensional de Sitter metrics, which correspond individually to left-movers and right-movers in the CFT.\footnote{Since we have used flat space CFT coordinates, this metric describes kinematic space for a Poincar\'{e} patch of AdS and covers only half the kinematic space of global AdS.} Of course, this decomposition reflects the factorization of the two-dimensional conformal symmetry.

\begin{figure}[t]
\centering
\includegraphics[width=.8\textwidth]{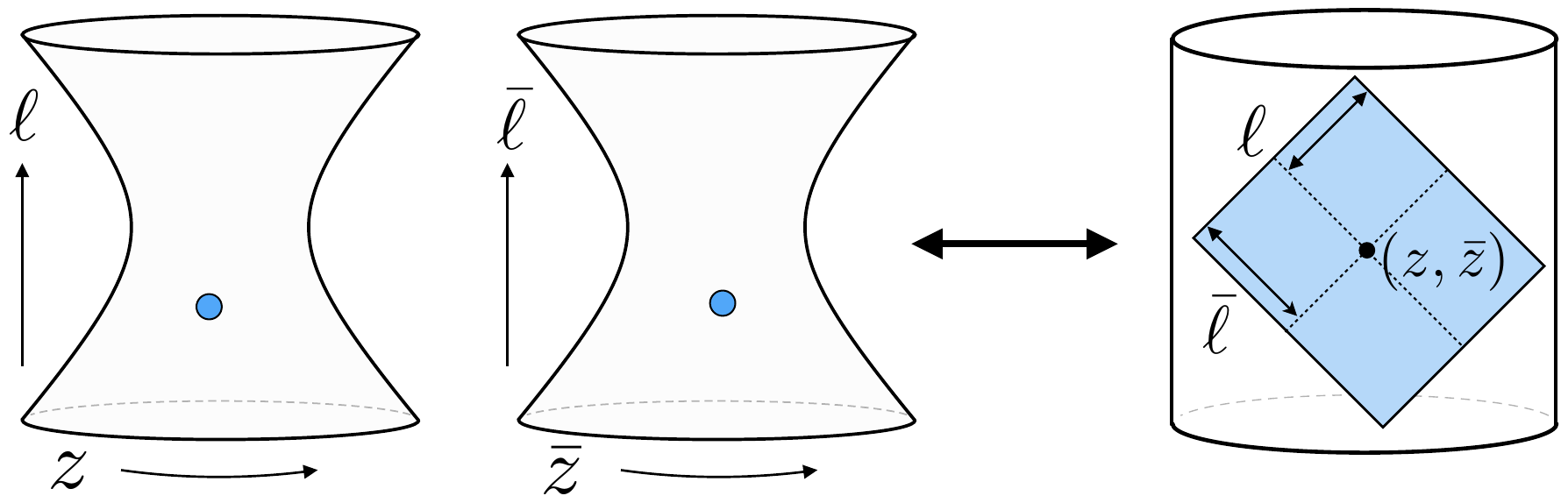}
\caption{The coordinates (\ref{eq:defkscoords}) of kinematic space represent the center position and the half-width of the causal diamond in the left-moving light-like coordinate; analogous relations define $\bar{z}$ and $\bar{\ell}$.}
\Label{fig:ks-coordinates}
\end{figure}

We may re-cast each de Sitter component of (\ref{eq:ksads3}) in more familiar, ``co-moving'' coordinates
\begin{equation}
\ell=\frac{z_1-z_2}{2} \qquad {\rm and} \qquad z=\frac{z_1+z_2}{2}
\label{eq:defkscoords}
\end{equation}
and likewise for the right-movers; see Fig.~\ref{fig:ks-coordinates}. Thus, $\ell$ is half the left-moving separation between $x_1$ and $x_2$ while $z$ is their average left-moving location. This coordinate change brings eq.~(\ref{eq:ksads3}) to the form:
\begin{equation}
ds^{2} = \frac{1}{2}\left[\frac{-d\ell^{2}+dz^2}{\ell^{2}}+\frac{-d\bar{\ell}^{2}+d\bar{z}^2}{\bar{\ell}^{2}}\right]
\label{eq:kscoords}
\end{equation}
We can restrict to an $\mathbb{H}^2$ slice of $\rm{AdS}_3$ by setting $\ell = \bar\ell, z = \bar z$; this reveals the single $\rm{dS}_2$ kinematic space discussed in \cite{Czech:2014ppa,Czech:2015qta}.

\paragraph{Causal structure} How can we understand the causal structure of each de Sitter component? For definiteness, let us focus on the $z$ (left-moving) de Sitter space. Consider two causal diamonds with left-moving coordinates $(z_1, z_2)$ and $(w_1, w_2)$. We temporarily ignore the right-moving sizes of the causal diamonds, effectively working with their projections onto the left-moving axis. When $(z_1, z_2) \subset (w_1, w_2)$ as intervals on the real line, $(z_1, z_2)$ causally precedes $(w_1, w_2)$ in the left-moving de Sitter component. If neither interval contains the other, the two intervals are not causally related.

The same rules apply to the right-moving de Sitter component. In the end, the causal structure of the AdS$_3$ kinematic space contains several distinct options, which are illustrated in Fig.~\ref{fig:ks-causal}. Unlike a generic space of $(2,2)$ signature, these options are well-defined because the kinematic space decomposes into two independent components.  It is useful conceptually to combine these two causal structures into an overarching structure, where $(z_1,z_2)$ precedes $(w_1,w_2)$ if and only if the corresponding causal diamonds satisfy $\diamond_{z} \subset \diamond_w$.

\begin{figure}[t]
\centering
\includegraphics[height=.4\textwidth]{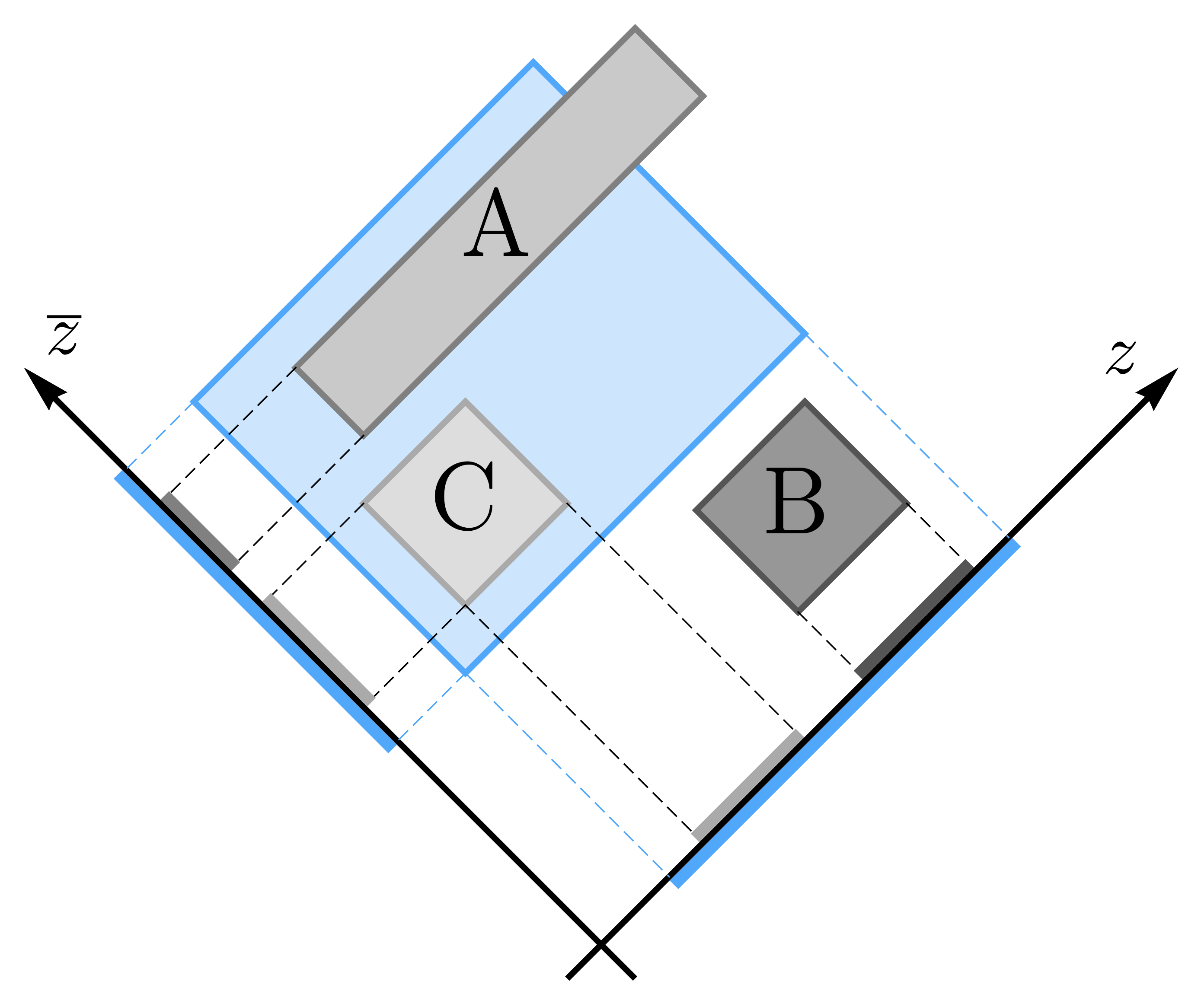}
\caption{When the left-moving projection of one diamond contains the left-moving projection of another, they are time-like separated in the $z$ (left-moving) factor of kinematic space; similar relations apply in the $\bar{z}$ (right-moving) component. This leads to several possible causal relations between two intervals, e.g. the big blue causal diamond is in the $z$-future of diamonds A and C and in the $\bar{z}$-future of diamonds B and C.  In the overarching causal structure, the blue diamond is preceded by C, but not related to A or B.}
\Label{fig:ks-causal}
\end{figure}

\section{OPE Blocks}
\label{opeblocks}

\subsection{OPE Kinematics}
In conformal field theories, quasiprimaries $\mathcal{O}_{i}\left(0\right)$
and their descendants $\partial_{\mu}\partial_{\nu}\cdots O_{i}\left(0\right)$
form a complete basis of operators. Any operator in the theory can be expanded in this basis as long as other operator insertions are sufficiently far away.

Consider a product of two separated scalar operators
$\mathcal{O}_{i}\left(x\right)\mathcal{O}_{j}\left(0\right)$ with conformal weights $\Delta_i$ and $\Delta_j$. Expanding it in a local basis centered at 0 gives\footnote{Conventionally, we expand in the operator basis at the location of
the second operator (0 above), but generally any point can be used
as long as it is sufficiently far from any other operator insertion.}
\begin{eqnarray}
\mathcal{O}_{i}\left(x\right)\mathcal{O}_{j}\left(0\right) & = & \sum_{k}C_{ijk}\left|x\right|^{\Delta_{k}-\Delta_{i}-\Delta_{j}}\big(1+b_{1}\,x^{\mu}\partial_{\mu}+b_{2}\,x^{\mu}x^{\nu}\partial_{\mu}\partial_{\nu}+\ldots\big)\mathcal{O}_{k}\left(0\right),\label{eq:local-ope}
\end{eqnarray}
where the sum ranges over quasiprimaries with definite scaling dimensions $\Delta_k$. The constants $C_{ijk}$ are the only 
theory-dependent, dynamical parameters in this expression; they are the OPE coefficients. Importantly, the coefficients $b_n$ depend only on the dimensions $\Delta_i,\Delta_j,\Delta_k$ and are determined entirely by the kinematics of conformal symmetry \cite{Belavin:1984vu}.

We wish to absorb the series of descendants appearing in every element of the sum (\ref{eq:local-ope}) in the definition of a new operator $\mathcal{B}_{k}^{ij}\left(x_1,x_2\right)$:
\begin{eqnarray}
\mathcal{O}_{i}\left(x_1\right)\mathcal{O}_{j}\left(x_2\right) & = & \left|x_1-x_2\right|^{-\Delta_{i}-\Delta_{j}}\sum_{k}C_{ijk}\mathcal{B}_{k}^{ij}\left(x_1,x_2\right),\label{eq:ope-block-definition}
\end{eqnarray}
where we have now generalized to arbitrary operator locations $x_1$ and $x_2$.\footnote{Formally, the OPE does not produce an operator, but a class of operators that act equivalently in a suitable space of states. For some readers, this may sound similar to the story of bulk operators and error correction \cite{Almheiri:2014lwa}. Later, we will choose particularly useful representatives of this class.} We
will call the operators $\mathcal{B}_{k}^{ij}\left(x_1,x_2\right)$ 
\emph{OPE blocks} because they are the building blocks of the operator product expansion. 

OPE blocks are \emph{non-local} operators in the CFT, but they have a functional dependence on pairs of CFT points. For this reason, it is natural to think of them as fields on kinematic space. We will often refer to OPE blocks as bi-locals to emphasize their dependence on pairs of CFT locations, though the reader should bear in mind the above caveat in this terminology.

\paragraph{Transformation properties of OPE blocks}
We now turn our attention to the representation theory of the OPE
blocks $\mathcal{B}_{k}^{ij}\left(x_1,x_2\right)$. Recall that under a conformal transformation $x\rightarrow x^{\prime}$, a spin-0 local operator transforms as 
\begin{equation}
\mathcal{O}_{i}\left(x\right)\rightarrow\Omega\left(x^{\prime}\right)^{\Delta}\mathcal{O}_{i}\left(x^{\prime}\right),\label{eq:local-operator-transformation}
\end{equation}
where the position-dependent rescaling $\Omega$ is: 
\begin{equation}
\Omega\left(x^{\prime}\right)=\det\left(\frac{\partial x^{\prime\mu}}{\partial x^{\nu}}\right).
\end{equation}
Moreover, the proper distance between two CFT points transforms as: 
\begin{equation}
\left(x_1-x_2\right)^{2}=\frac{\left(x_1^{\prime}-x_2^{\prime}\right)^{2}}{\Omega\left(x_1^{\prime}\right)\Omega\left(x_2^{\prime}\right)}.
\end{equation}
Combining these well-known facts, we can readily derive the transformation properties of the OPE block:
\begin{equation}
\mathcal{B}_{k}^{ij}\left(x_1,x_2\right)\rightarrow\left(\frac{\Omega\left(x_1^{\prime}\right)}{\Omega\left(x_2^{\prime}\right)}\right)^{\left(\Delta_{i}-\Delta_{j}\right)/2}\mathcal{B}_{k}^{ij}\left(x_1^{\prime},x_2^{\prime}\right).\label{eq:projected-ope-frame-dependence}
\end{equation}
Specializing to the case $\Delta_{i}=\Delta_{j}$, the OPE block transforms simply as
\begin{equation}
\mathcal{B}_{k}\left(x_1,x_2\right)\rightarrow\mathcal{B}_{k}\left(x_1^{\prime},x_2^{\prime}\right), \label{eq:blockscalarfield}
\end{equation}
where we drop the dependence on the external operator dimensions to reduce clutter. This simplification of notation is further justified since, as we will see shortly, the form of OPE blocks is in fact insensitive to the external weights when $\Delta_i=\Delta_j$.\footnote{In general, the scalar OPE blocks in fact depend only on $\Delta_i-\Delta_j$.}
We have already suggested that the bi-local operator $\mathcal{B}_k(x_1,x_2)$ is a natural kinematic space object. The transformation law (\ref{eq:blockscalarfield}) means that we should identify it with a scalar field. This observation will give us a lot of mileage in the upcoming sections.

In the case of products of local scalar operators with unequal weights $(\Delta_i\neq \Delta_j)$ OPE blocks also turn out to be scalar fields in kinematic space. The only difference is that the kinematic scalar $\mathcal{B}^{ij}_k(x_1,x_2)$ is charged under a decompactified global $U(1)$ symmetry, which is related to special conformal transformations.

\subsection{OPE Blocks as Kinematic Space Fields}
Our next goal is to prove that OPE blocks obey the Klein-Gordon equation in kinematic space. To do so, we need one additional property of $\mathcal{B}_k(x,y)$: that they are eigenoperators of the conformal Casimir. We will then recognize that the Casimir eigenvalue equation is the Klein-Gordon equation in metric~(\ref{eq:kscoords}).

Let $L_{0,\pm 1}$ and $\bar{L}_{0,\pm 1}$ be the standard generators of the global conformal group $\mathrm{SO}\left(2,2\right)$. Their algebra is represented on conformal fields $\mathcal{O}_{k}\left(x\right)$ by appropriate differential operators $\mathcal{L}_{\left(k\right)AB}$ via $\left[L_{AB},\mathcal{O}_{k}\left(x\right)\right]=\mathcal{L}_{\left(k\right)AB}\mathcal{O}_{k}\left(x\right)$. 

Irreducible representations of the conformal group are classified by their eigenvalues under the Casimir operator\footnote{Note that this convention for the Casimir differs by a factor of two from the usual 2D CFT convention; this is useful for generalizing to higher dimensions.}:
\begin{equation}
L^2 = L_{AB} L^{AB} \equiv \left( -2 L_{0}^{2}+L_{1}L_{-1}+L_{-1}L_{1} \right) +\left(L\rightarrow\bar{L}\right).
\label{eq:ccasimir}
\end{equation}
In particular, all descendants of a quasiprimary operator $\mathcal{O}_{k}$ live in the same eigenspace as operator~(\ref{eq:ccasimir}) and satisfy the same eigenvalue equation:
\myeq{\left[L^{2},\partial_{\mu_1} \ldots \partial_{\mu_p}\mathcal{O}_{k}\left(x\right)\right]= 
\mathcal{L}_{\left(k\right)AB}\mathcal{L}_{\left(k\right)}^{AB}\, \partial_{\mu_1} \ldots \partial_{\mu_p} \mathcal{O}_k(x)
=C_{k}\,\partial_{\mu_1} \ldots \partial_{\mu_p}\mathcal{O}_{k}\left(x\right)}
The eigenvalue is:
\begin{equation}
C_{k}=-\Delta_k\left(\Delta_k-d\right)-\ell_k\left(\ell_k+d-2\right),
\label{eq:cceigenvalue}
\end{equation}
where $\Delta_k$ and $\ell_k$ denote the scaling dimension and spin of the quasiprimary $\mathcal{O}_{k}$, and where $d=2$ here.

Every OPE block is a linear combination of a single quasiprimary operator and its descendants. Therefore, the operator $\mathcal{B}_k(x,y)$ is also an eigenvector of the conformal Casimir and obeys:
\begin{equation}
\left[L^{2},\mathcal{B}_{k}\left(x_1,x_2\right)\right]=
C_k\,\mathcal{B}_{k}\left(x_1,x_2\right)=
\mathcal{L}_{\left(B\right)}^{2}\mathcal{B}_{k}\left(x_1,x_2\right)
\label{eq:beigenvector}
\end{equation}
In the second equality, we again represent the Casimir as some differential operator $\mathcal{L}_{\left(B\right)}^{2}$, which now acts on $x_1$ and $x_2$. To identify that representation, recall that $\mathcal{B}_k(x_1,x_2)$ transforms as a scalar function of both arguments; see eq.~(\ref{eq:blockscalarfield}). Therefore the appropriate representation can be built from two local field representations with $\Delta=0$: \begin{equation}
\mathcal{L}_{\left(B\right)}^{2}=\left(\mathcal{L}_{\left(0,x_1\right)}+\mathcal{L}_{\left(0,x_2\right)}\right)^{2} = \left(\mathcal{L}_{\left(0,x_1\right)AB}+\mathcal{L}_{\left(0,x_2\right)AB}\right) \left(\mathcal{L}_{\left(0,x_1\right)}^{AB}+\mathcal{L}_{\left(0,x_2\right)}^{AB}\right)
\label{eq:repb}
\end{equation}
Expressing $x_1$ in light-like coordinates $z_1$ and $\bar{z}_1$, the representation $\mathcal{L}_{(0,x_1)0,\pm1}$ of $L_{0,\pm 1}$ takes the form: 
\begin{equation}
\mathcal{L}_{(0,x_1)0}= - z_1\,\partial_{z_1}
\qquad {\rm and} \qquad
\mathcal{L}_{(0,x_1)1}= -i z_1^{2}\,\partial_{z_1}
\qquad {\rm and} \qquad 
\mathcal{L}_{(0,x_2)-1}= i \partial_{z_1},
\label{eq:rep0}
\end{equation}
with similar formulas for the right-movers and for $x_2$. Using eqs.~(\ref{eq:rep0}) and (\ref{eq:ccasimir}), we therefore obtain:
\begin{equation}
\mathcal{L}_{\left(B\right)}^{2} 
=  2\left[\square_{\mathrm{dS}_{2}}+\square_{\overline{\mathrm{dS}}_{2}}\right]
= 2\left[\ell^{2}\left(-\partial_{\ell}^{2}+\partial_{z}^{2}\right)+\bar{\ell}^{2}\left(-\partial_{\bar{\ell}}^{2}+\partial_{\bar{z}}^{2}\right)\right]
\end{equation}
This is the Laplacian in metric~(\ref{eq:kscoords}). On the right, we traded the coordinates $z_1$ and $z_2$ for $\ell$ and $z$, which were defined in eq.~(\ref{eq:defkscoords}).  The appearance of the kinematic space Laplacian comes from the fact that kinematic space is a homogeneous space of the conformal group; see Appendix \ref{homogeneous} for details.

If $\mathcal{L}_{(B)}^2$ is the Laplacian then eq.~(\ref{eq:beigenvector}) is the Klein-Gordon equation:
\begin{equation}
2 \left(\square_{\mathrm{dS}_{2}}+\square_{\overline{\mathrm{dS}}_{2}}\right)  \mathcal{B}_{k}\left(x_1,x_2\right) = C_k \mathcal{B}_{k}\left(x_1,x_2\right)
\label{eq:blockKG}
\end{equation}
The mass-squared term is the constant $C_k$ defined in eq.~(\ref{eq:cceigenvalue}). It is negative, so $\mathcal{B}_{k}\left(x_1,x_2\right)$ is a tachyon in kinematic space. We will see shortly, however, that this does not lead to inconsistencies.

In fact, the two-dimensional conformal group has another quadratic Casimir operator which characterizes the spin of a representation:
\begin{equation}
S=\left(-2 L_{0}^{2}+L_{1}L_{-1}+L_{-1}L_{1}\right)-\left(L\rightarrow\bar{L}\right) = 2 \ell (\Delta - 1)
\end{equation}
As is easy to guess, its representation as a differential operator on bi-locals is $2\left(\square_{{\rm dS}_{2}}-\square_{\overline{\mathrm{dS}}_{2}}\right)$. This gives us another differential equation obeyed by $\mathcal{B}_k(x_1,x_2)$:
\begin{equation}
2 \left( \square_{{\mathrm{dS}}_{2}}-\square_{\overline{\mathrm{dS}}_{2}} \right)\mathcal{B}_{k}\left(x_1,x_2\right)=2\ell \left(\Delta -1 \right) \mathcal{B}_k(x_1,x_2).
\label{eq:blockconstraint}
\end{equation}
Eqs.~(\ref{eq:blockKG}) and (\ref{eq:blockconstraint}) are the two kinematic ``equations of motion'' for the OPE block. Note that eq.~(\ref{eq:blockconstraint}) decouples the ``time-evolution'' of the OPE block in the left-moving and right-moving sectors. In other words, finding the OPE block requires solving two 1+1-dimensional problems rather than a single 2+2-dimensional problem. To select the right solution, we must supplant the Klein-Gordon equations with appropriate boundary conditions.

\begin{figure}
\centering{\includegraphics[width=0.8\textwidth]{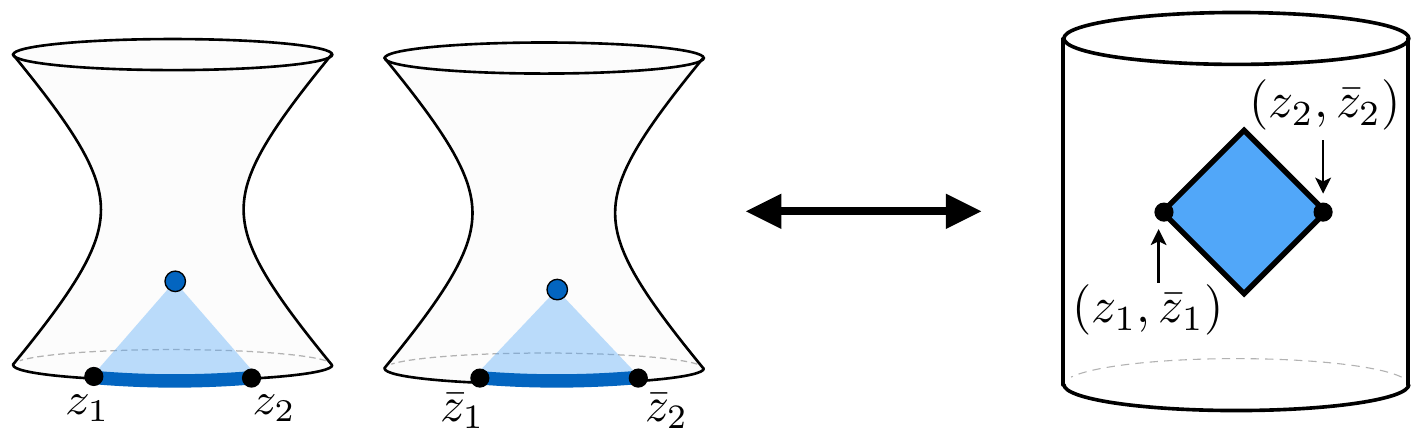}
\caption{Causality in each de Sitter component of kinematic space means that the OPE block at $(z_1, \bar{z}_1, z_2, \bar{z}_2)$ depends only on the initial data between $z_1$ and $z_2$ in the first component and between $\bar{z}_1$ and $\bar{z}_2$ in the second component. These loci span the CFT causal diamond with corners at $x_1$ and $x_2$. 
\label{fig:causal-past-of-ks}}}
\end{figure}

\subsection{Smeared Representation of OPE Blocks}
We will have a well-defined Cauchy problem if we specify a set of initial conditions on each de Sitter component of kinematic space. In coordinates (\ref{eq:defkscoords}), the asymptotic past is reached when we send $\ell, \bar{\ell}$ to 0. In this limit $x_1$ and $x_2$, the two CFT locations parametrizing $\mathcal{B}_{k}\left(x_1,x_2\right)$, approach one another and the bi-local reduces to a local operator! Because the coefficients of descendants are suppressed by higher powers of $|x_1-x_2|$, the correct initial condition comes from the leading-order contribution to Eq. (\ref{eq:local-ope}),
\begin{equation}
\lim_{x_2\to x_1} \mathcal{B}_{k}\left(x_1,x_2\right)=
\left(z_1-z_2\right)^{h_{k}}\left(\bar{z}_1-\bar{z}_2\right)^{\bar{h}_{k}}\mathcal{O}_{k}\left(x_1\right),
\label{eq:ope-block-boundary-conditions}
\end{equation}
where $\mathcal{O}_{k}(x)$ is the quasi-primary that labels the OPE block. In this expression, we used the standard left/right-moving conformal weights: $h_k=\frac{1}{2}\left(\Delta_k+\ell_k\right)$ and $\bar{h}_k=\frac{1}{2}\left(\Delta_k-\ell_k\right)$.

\begin{figure}[t!]
\noindent \centering{}\includegraphics[width=0.5\textwidth]{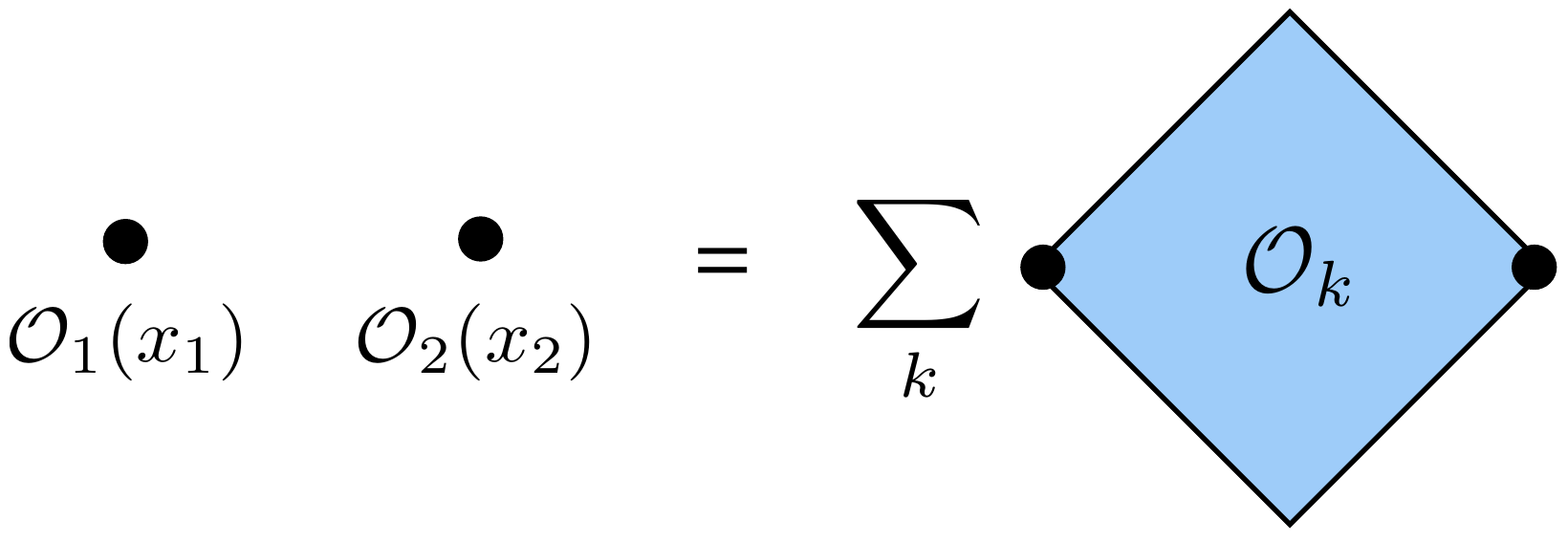}\caption{A product of scalar $\mathrm{CFT}_{2}$ operators inserted at $x_1$ and $x_2$ can be expanded
in terms of OPE blocks, which consist of primary operators smeared over the causal diamond $\diamond_{12}$.
\label{fig:bilocaltosmeared}}
\end{figure}

All that remains is to write down a kinematic boundary-to-bulk propagator. The decoupling of the left and right-movers means that it will be a product of two respective propagators. This gives the following schematic form of the OPE block:
\begin{equation}
\mathcal{B}_{k}\left(x_1,x_2\right)=
\int dw\, G_{k}\left(w;\,z_1,z_2\right)  
\int d\bar{w}\, 
\bar{G}_{k}\left(\bar{w};\,\bar{z}_1,\bar{z}_2\right) 
\mathcal{O}_{k}\left(w,\bar{w}\right)
\label{eq:cft2-smeared-ope}
\end{equation}
If we choose $G_k(w;\,z_1, z_2)$ to be the advanced propagator, the solution will respect causality in kinematic space. This choice means that $G_k(w;\,z_1, z_2) = 0$ unless $z_1 < w < z_2$, so that the $w$-integral in eq.~(\ref{eq:cft2-smeared-ope}) extends from $z_1$ to $z_2$; see Fig.~\ref{fig:causal-past-of-ks}. Taking into account the analogous limits for the $\bar{w}$-integral, we conclude that the integrals in eq.~(\ref{eq:cft2-smeared-ope}) cover ${\diamond_{12}}$, the causal diamond defined by $x_1$ and $x_2$. Most importantly, the choice of advanced propagator allows us to impose the boundary conditions (\ref{eq:ope-block-boundary-conditions}): as $x_2$ approaches $x_1$, the diamond $\diamond_{12}$ (and hence the support of the advanced propagator) covers a small neighborhood of $x_1$, and the resultant block is localized at that point.
 The explicit form of the advanced propagator is:
\begin{equation}
G_k(w;\,z_1, z_2) \propto \left(\frac{(w-z_1)(z_2-w)}{z_2-z_1}\right)^{h_{k}-1}.
\end{equation}
Collecting these facts and fixing the normalization from eq.~(\ref{eq:ope-block-boundary-conditions}), we find the smeared form of the OPE block:
\begin{equation}
\mathcal{B}_{k}\left(x_1,x_2\right)\!=\!
\frac{\Gamma\left(2h_k\right)\Gamma(2\bar{h}_k)}{\Gamma(h_k)^2\, \Gamma(\bar{h}_k)^2}\!\!
\int_{\diamond_{12}}\!\!\!\! dw\, d\bar{w}\!
\left(\frac{(w-z_1)(z_2-w)}{z_2-z_1}\right)^{\! h_{k}-1}  \!\!
\left(\frac{(\bar{w}-\bar{z}_1)(\bar{z}_2-\bar{w})}{\bar{z}_2-\bar{z}_1}\right)^{\! \bar{h}_{k}-1}\!\!
\mathcal{O}_{k}\left(w,\bar{w}\right)
\label{eq:smeared}
\end{equation}
In this way, the product of scalar operators
$\mathcal{O}_{i}\left(x_1\right)\mathcal{O}_{j}\left(x_2\right)$ can
be expanded in terms of quasiprimary operators that are \emph{smeared}
over the causal diamond $\diamond_{12}$; see Fig.~\ref{fig:bilocaltosmeared}.

Formula (\ref{eq:smeared}) can also be derived in a different way related to the shadow operator formalism. In that language, the OPE block becomes:
\begin{equation}
\mathcal{B}_{k}^{ij}\left(x_1,x_2\right) \propto
\int d^{d}z\,\left|x_1-x_2\right|^{\Delta_{i}+\Delta_{j}}\left\langle \mathcal{O}_{i}\left(x_1\right)\mathcal{O}_{j}\left(x_2\right)\tilde{\mathcal{O}}_{k\,\mu\nu\ldots}\left(z\right)\right\rangle \,\mathcal{O}_{k}^{\mu\nu\ldots}\left(z\right)
\end{equation}
We explain the shadow operator method, which is better suited to higher-dimensional generalizations, in Appendix~\ref{app:shadows}.

\section{Geodesic Operators}
\label{gops}

We discussed OPE blocks in the hope that they form an ideal holographic operator basis---as we characterized them in the introduction. To realize this hope, OPE blocks must have a natural bulk interpretation. The fact that OPE blocks live in kinematic space---the space of bulk geodesics---suggests a guess for their holographic dual. This guess is the X-ray transform, an integral of an (operator-valued) function along a geodesic. In the holographic context, the X-ray transform has appeared e.g. in \cite{Lin:2014hva}.

In this section, we confirm that the correspondence between OPE blocks and \emph{geodesic operators} is correct.

\subsection{A Brief Introduction to X-ray Transforms}

Integral geometry supplies us with canonical maps from local functions defined on a manifold $M$ to functions on the space of totally geodesic submanifolds of dimension $k$ \cite{Helgason1999,Helgason2011}. These maps, obtained by integrating the function over a submanifold, are known in general as \emph{Radon transforms}. For $M=\mathrm{AdS}_{d+1}$, we will be particularly interested in the cases of $k=1$ and $k=d-1$, which correspond to geodesics and codimension-2 minimal surfaces, respectively.

For now, as we discuss AdS$_3$, there is only one transform to consider: the geodesic Radon transform or \emph{X-ray transform}. 
Thus, consider the interpretation of kinematic space $\mathcal{K}(M)$ as the space of boundary-anchored spacelike geodesics in $M$. Given a function $f:M \rightarrow\mathbb{R}$, we can define its X-ray transform $Rf:\mathcal{K}\left(M\right)\rightarrow\mathbb{R}$
as
\begin{equation}
Rf\left(\gamma\right)=\int_{\gamma}ds\,f\left(x\right).\label{eq:xray-definition}
\end{equation}
In other words, $Rf\left(\gamma\right)$ is the integral
of $f$ over the geodesic $\gamma$, weighted by its proper length.

An important property of the X-ray transform---which we will exploit in this paper---is that it is known to be invertible when $M$ is either hyperbolic space or flat space of any dimension.\footnote{More generally, there are inversion formulas for Radon transforms on the totally geodesic submanifolds of arbitrary dimension \cite{Helgason1999,Helgason2011,Rubin2002206}.} Given only knowledge
of $Rf$ we can recover the function
$f$ on the entire manifold $M$ by using an appropriate inversion formula. 
We discuss this inversion formula in more detail in Sec.~\ref{localbulkops}.

\subsection{Kinematic Operators from Bulk Fields}

In AdS/CFT, the bulk theory is described at low energies by an effective
field theory. The relevant degrees of freedom are the propagating
excitations of a corresponding field, which can be locally created by field operators such as $\phi\left(x\right)$ for a spin-0 particle.

Let us consider the case of a free scalar in ${\rm AdS}_{d+1}$ with mass $m^{2}$. Its propagation is described by the Klein-Gordon equation:
\begin{equation}
\left(\square_{{\rm AdS}}-m^{2}\right)\phi\left(x\right)=0.\label{eq:AdSKG}
\end{equation}
Since we are interested in the bulk operator $\phi$ which creates quantum states of finite norm, only the regular, normalizable solutions of (\ref{eq:AdSKG}) describe the relevant field modes. Moreover, according to the standard AdS/CFT dictionary \cite{wittenadscft}, the operator $\phi$ is dual
to a  single-trace primary CFT operator of spin $\ell=0$ whose weight $m^{2}=\Delta\left(\Delta-d\right)$ is determined by the conformal Casimir (\ref{eq:cceigenvalue}). In the extrapolate version of the dictionary, the two operators are related by
\begin{equation}
\phi\left(z\rightarrow0,x\right)\sim z^{\Delta}\mathcal{O}_{\Delta}\left(x\right)\label{eq:AdSbc}
\end{equation}
in the absence of sources.

Using our knowledge of the X-ray transform introduced in
the previous section, we can map the local operator basis $\phi(x)$ to operators
$\tilde{\phi}\left(\gamma\right)=R\phi\left(\gamma\right)$ on kinematic space.

\paragraph{Intertwinement of the Laplacian}

A natural question to ask is whether the equation of motion satisfied
by a field $\phi$ in AdS implies an equation of motion for its X-ray transform
$\tilde{\phi}$. In fact, we will see that this is the case. The equation of motion for the geodesic integral of a field follows from the \emph{intertwining} property of the X-ray transform: the kinematic space Laplacian acting on the X-ray transform of $f(x)$ is equal to the X-ray transform of the AdS Laplacian acting on $f(x)$.

We will now prove this property in the simplest way available. The key fact is that a shift of the function $f$ by some isometry of ${\rm AdS}$, can be compensated for by a corresponding shift of the
function $Rf$.

\begin{figure}
\centering{\includegraphics[width=0.8\textwidth]{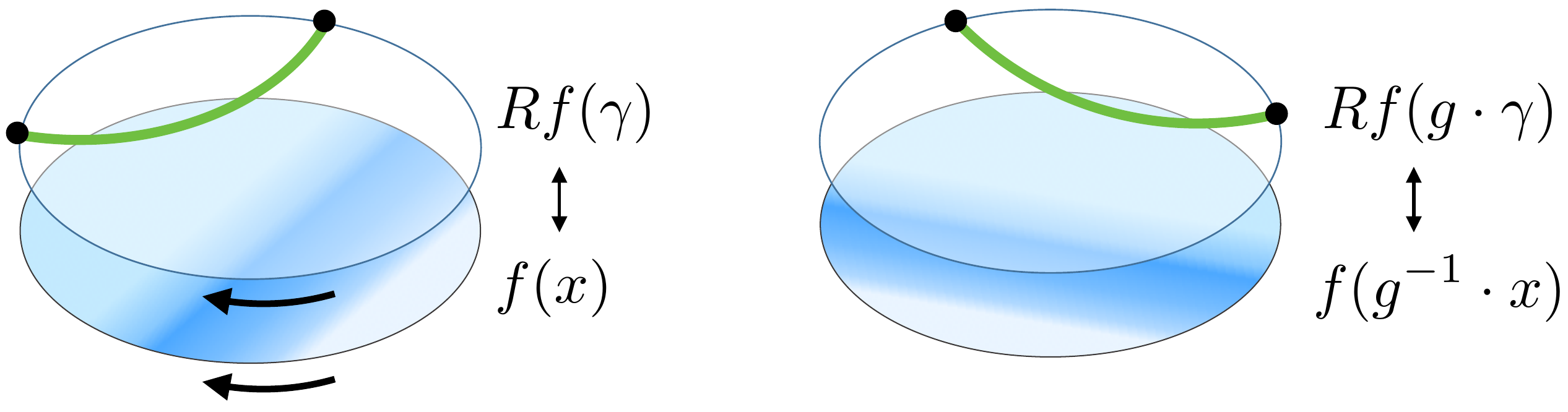}
\caption{A shift in the field configuration by an AdS isometry can be compensated by a corresponding shift in the X-ray transform.  This allows us to derive an intertwining relation (\ref{generator-intertwinement}) between differential operators acting on AdS and kinematic space fields. \label{fig:intertwinement}}}
\end{figure}

Consider the X-ray transform $Rf\left(\gamma\right)$ of some function $f\left(x\right)$ and let $g\in{\rm SO}\left(2,2\right)$ be an isometry of ${\rm AdS}_{3}$.\footnote{Though phrased in terms of $\rm{AdS}_3$, this proof is valid for any dimension of AdS and indeed for any pair of homogeneous spaces.} This group element acts on the manifolds ${\rm AdS}_{3}$ and $\mathcal{K}\left({\rm AdS}_{3}\right)$
in the obvious way. Consider now the function $f^{\prime}\left(x\right)=f\left(g^{-1}\cdot x\right)$,
which is just a shifted version of $f$. We can evaluate the X-ray
transform of $f^{\prime}$ by a shift of the integration path:
\begin{eqnarray}
Rf^{\prime}\left(\gamma\right) & = & \int_{\gamma}f\left(g^{-1}\cdot x\right)ds\nonumber \\
 & = & \int_{g\cdot\gamma}f\left(x\right)ds=Rf\left(g\cdot\gamma\right) \label{shifting-intertwinement}
\end{eqnarray}
Hence, the shift of $f$ can be compensated for by a corresponding
shift in $Rf$; see Fig.~\ref{fig:intertwinement}.
Now, let $g$ be a group element near the identity. Then, we can write
\begin{eqnarray}
f^\prime\left(x\right) & = & \left(1-\omega^{AB}L_{AB}^{\left(x\right)}\right)f\left(x\right)\nonumber \\
Rf\left(g\cdot\gamma\right) & = & \left(1+\omega^{AB}L_{AB}^{\left(\gamma\right)}\right)f\left(\gamma\right)
\end{eqnarray}
where $L_{AB}^{\left(x\right)},L_{AB}^{\left(\gamma\right)}$ are
the ${\rm AdS}_{3}$ and kinematic space scalar field representations
of $\mathfrak{so}\left(2,2\right)$, respectively, and $\omega^{AB}$ parametrize the choice of $g$.

Using the equality $Rf^\prime \left(\gamma\right) = Rf(g\cdot\gamma)$, we find that the differential operators  $L_{AB}^{\left(x\right)}$ and $L_{AB}^{\left(\gamma\right)}$ \emph{intertwine} under the X-ray transform: 
\begin{equation}
L_{AB}^{\left(\gamma\right)}Rf= -RL_{AB}^{\left(x\right)}f. \label{generator-intertwinement}
\end{equation}
Applying this relationship twice, we can construct the Casimir operator
$L^{\left(x\right)2}Rf=RL^{\left(\gamma\right)2}f$.
Since the Casimir operators $L^{\left(x\right)2}$ and $L^{\left(\gamma\right)2}$
are represented by the Laplace operators $-\square_{{\rm AdS_3}}$ and
$\square_{\mathcal{K}}=2 \left(\square_{\mathrm{dS}_{2}}+\square_{\overline{\mathrm{dS}}_{2}}\right)$ respectively (see Appendix \ref{homogeneous}), we find
the \emph{intertwining property} of the Laplacian:
\begin{equation}
2 \left(\square_{\mathrm{dS}_{2}}+\square_{\overline{\mathrm{dS}}_{2}}\right)Rf=-R\square_{{\rm AdS}}f. \label{ads3-intertwinement-laplacian}
\end{equation}
In other words, the AdS Laplacian \emph{intertwines} with the kinematic
space Laplacian.

Consequently, the X-ray transform $\tilde\phi = R\phi$ of a free scalar field $\phi$ of mass $m^{2}$
defines a free field of mass $-m^{2}$ propagating on the kinematic geometry:
\begin{equation}
\left(\square_{{\rm AdS}}-m^{2}\right)\phi\left(x\right)=0\quad\implies\quad\left(2 \left(\square_{\mathrm{dS}_{2}}+\square_{\overline{\mathrm{dS}}_{2}}\right)+m^{2}\right)\tilde{\phi}\left(\gamma\right)=0. \label{eq:KGforXray}
\end{equation}
By referring to eq.~(\ref{eq:blockKG}), we see that this is precisely
the same equation as is obeyed by the CFT dual
of $\tilde\phi\left(\gamma\right)$---the OPE block $\mathcal{B}_{\Delta}$
of the primary associated with $\phi(x)$.

\begin{figure}
\centering{\includegraphics[height=0.25\textwidth]{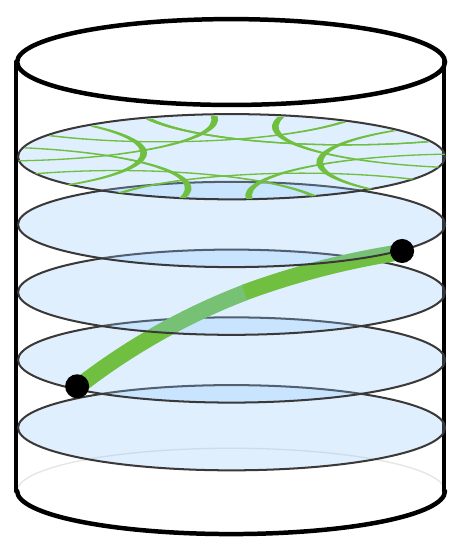}
\caption{The X-ray transform of a function on $\rm{AdS}_3$ obeys a constraint equation.  Given access only to geodesics living on a given $\mathbb{H}_2$ slice of $\rm{AdS}_3$, the X-ray transform can be inverted on that slice.  Thus, we only need access to the unboosted geodesics to recover an entire function on AdS.  In this sense, the information in the boosted geodesics is redundantly encoded.	 \label{fig:johns-equations}}}
\end{figure}

\paragraph{Constraint equations}
The AdS$_{3}$ scalar field $\phi$ is a function of $3$ spacetime coordinates. In mapping this field to kinematic space via the X-ray transform, we obtained a function of $4$ coordinates that parametrize the boundary locations of the geodesic endpoints. In other words, the X-ray transform introduces \emph{redundancies}: the geodesic integrals of a function on AdS$_3$ are an over-complete encoding of the said function (see Fig.~\ref{fig:johns-equations}).

An equivalent statement is that \emph{not} every function on the $4$-dimensional kinematic space can be understood as the X-ray transform of a function on AdS$_{3}$. One, therefore, needs to identify a set of \emph{constraint equations} that restrict kinematic functions $\tilde{\phi}(\gamma)$  to the ``physical subspace'' of consistent X-ray transforms. These extra equations ought to come from identities satisfied by our map to the space of geodesics.

The existence of non-trivial identities of X-ray transforms is a well-known fact in the mathematical literature and they were originally derived by Fritz John \cite{john1938} in the study of line integrals of functions in flat space. For AdS$_3$, we only have one equation which reads:
\myeq{2 \left(\Box_{\rm{dS}_2} -\Box_{\overline{\rm{dS}}_2}\right) Rf= 0 \label{eq:AdS3John}}
Eq.~(\ref{eq:AdS3John}) is, of course, identical to  (\ref{eq:blockconstraint}). That equation is satisfied by the OPE block of the dual CFT operator $\OO_{\Delta}$ as a dictated by  the second quadratic Casimir $S$ of SO(2,2).\footnote{Since we are here considering a scalar field, $\OO_{\Delta}$ has $\ell=0$.}  
The intertwining relation (\ref{generator-intertwinement}) guarantees that the differential representation of $S$ annihilates the X-ray transform.  It can also be verified explicitly using the $\rm{AdS}_3$ representation of the group generators from e.g. \cite{Balasubramanian:1998sn}.

As in the CFT discussion of Sec.~\ref{opeblocks}, the constraint equation can be combined with (\ref{eq:KGforXray}) to completely decouple the propagation of the geodesic operator on the two de Sitter components of kinematic space. This fact guarantees that the initial value problem for our system of differential equations is well posed.

\begin{figure}
\centering{\includegraphics[width=0.8\textwidth]{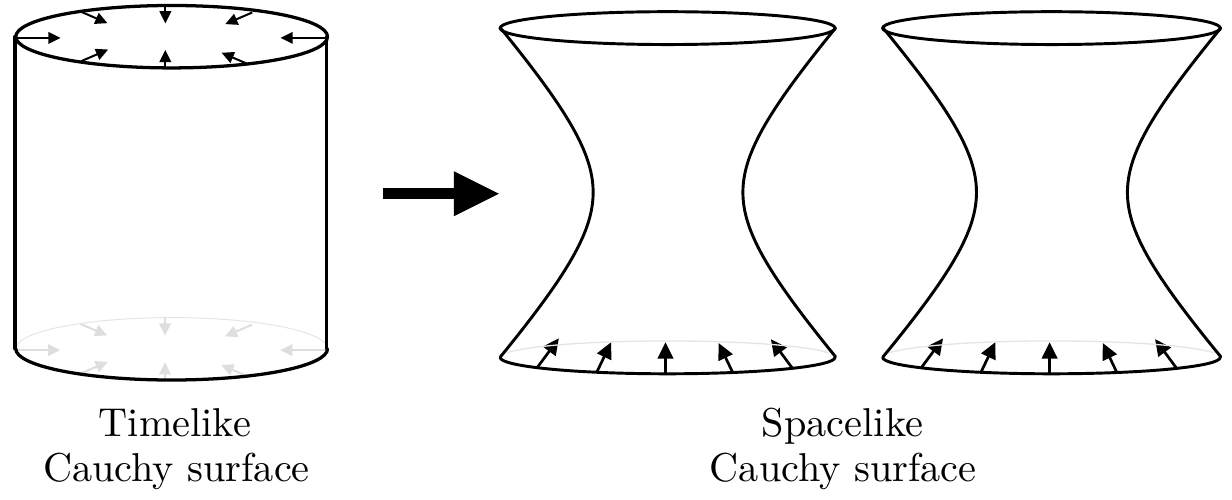}
\caption{The X-ray transform takes the non-standard bulk reconstruction problem and transforms it to a more standard Cauchy problem.  In particular, while the Cauchy data for the $\rm{AdS}_3$ reconstruction problem is given on a timelike surface, the corresponding data in kinematic space is on a spacelike surface.}
\label{fig:cauchy-problem}}
\end{figure}

\subsection{A Gauge-Invariant Holographic Dictionary}
The X-ray transform maps local field operators on AdS$_3$ to geodesic operators. The description of the latter as a local propagating excitation on kinematic space with equations of motion (\ref{eq:KGforXray}) and (\ref{eq:AdS3John}) will now allow us to connect geodesic bulk operators with OPE blocks on the boundary. In doing so, we take the first step towards a diffeomorphism invariant dictionary for AdS/CFT, valid at leading order in $N$. This will be one of the main results of this paper.

Both X-ray transforms of bulk fields and OPE blocks are defined via the same set of differential equations. Thus, proving they are equivalent operators amounts to merely verifying they also obey the same initial conditions. The asymptotic past of kinematic space, which we choose as a Cauchy surface for our initial value problem, is approached in the coincident limit of the bi-local: $x_2 \rightarrow x_1$ (see Fig.~\ref{fig:cauchy-problem}). As we have already discussed in Sec.~\ref{opeblocks}, OPE blocks in this limit behave like:
\myeq{\mathcal{B}_{k}(x_1,x_2) \underset{x_2 \rightarrow x_1}{\rightarrow} |x_2-x_1|^{\Delta_k} \OO_{k}(x_1)}

\begin{figure}
\centering{\includegraphics[width=0.8\textwidth]{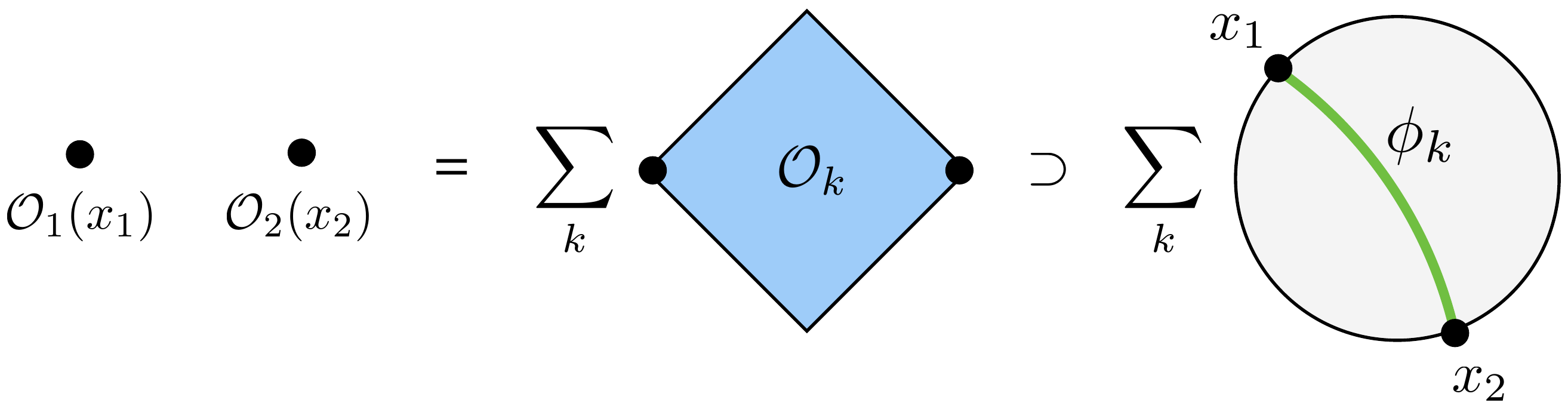}
\caption{The OPE block is represented on the boundary by a smeared diamond operator, and in the bulk (for low-dimension single-trace operators) by a geodesic operator.  \label{fig:ope-block-equals-geodesic}}}
\end{figure}

The boundary conditions for the X-ray transform are equally straightforward to derive. The geodesics anchored at the two boundary points defining the bi-local are contained in a neighborhood of the asymptotic boundary that can be made arbitrarily small as we send $x_2 \rightarrow x_1$. In this limit, the bulk field asymptotes to its dual primary operator in the CFT as in eq.~(\ref{eq:AdSbc}). Using the extrapolate dictionary we find that:
\myeq{\tilde{\phi}(x_1,x_2) \underset{x_2 \rightarrow x_1}{\rightarrow} \int ds \,\, z^{\Delta} \OO_{k}(x_1)\Big|_{\gamma} = \frac{\Gamma\left(\frac{\Delta}{2}\right)^2}{2\Gamma(\Delta)}\,\,|x_2-x_1|^{\Delta_k}\,\OO_{k}(x,\bar{x})}
We conclude that the OPE blocks in the CFT are dual to integrals of bulk local operators along geodesics; see Fig.~\ref{fig:ope-block-equals-geodesic}. Both objects behave as local excitations propagating in kinematic space:
\myeq{c_\Delta \,\,\mathcal{B}_{k}(x_1,x_2) =  \,\, \tilde{\phi}_k(\gamma_{12})= \int\limits_{\gamma_{12}} ds\,\, \phi\left(x\right). \label{kinematicdictionary}}
where $c_\Delta = \Gamma\left(\frac{\Delta}{2}\right)^2 / 2\Gamma(\Delta)$.  This completes the derivation of our gauge-invariant dictionary.

Thus far, we have treated the bulk field as freely propagating in AdS. This assumption is correct at leading order in $1/N$. However, bulk interactions will modify this dictionary at subleading orders. We comment on this briefly in the Discussion, saving a detailed analysis for a future publication. Nevertheless, even at leading order in $1/N$, the OPE block/geodesic operator equivalence wields considerable power. It reveals new insights to a number of holographic applications, to which we now turn.

\section{Construction of Bulk Local Operators}
\label{localbulkops}

Thus far we have explored bulk physics using non-local and diffeomorphism-invariant probes. 
Nevertheless, we would still like to understand the emergence of \emph{local} effective field theory in the gravitational background.

In Sec.~\ref{gops} we began our study of geodesic operators by starting with the real space geometry and integrating local operators along geodesics, exactly akin to how X-rays probe a density function in space. 
Inverting this process to determine the original local function is a well-studied problem (one necessary, for example, to display an intelligible CAT-scan image).
In this section, we will import these imaging techniques to reconstruct local operators. 
While the techniques we are discussing are quite general, we will focus on the example of a scalar field living in AdS$_3$ using the inverse X-ray transform on two-dimensional hyperbolic space.

We begin by discussing the inverse X-ray transform in hyperbolic space and then invert the transform for the analogous operator problem. 
The representation of the geodesic operators we use as input are exactly the OPE blocks of the CFT$_2$. 
The inversion formula gives a CFT representation for a local bulk operator at a point, which is defined invariantly on the boundary as the intersection locus of a family of geodesics.\footnote{Such a collection of geodesics, called a point-curve in \cite{Czech:2015qta}, will not define a point in an arbitrary background. It is a difficult problem to determine which families of geodesics intersect at a single point in a given geometry. An alternative---but not easier---way to specify a point involves its distances from all geodesics.}
We find that this representation of the bulk operator is exactly equivalent to the HKLL prescription \cite{Banks:1998dd,Balasubramanian:1998de,Bena:1999jv,Hamilton:2005ju,Hamilton:2006az,Heemskerk:2012np,Heemskerk:2012mn}: the geodesic operators deconstruct the HKLL representation into contributions of separate causal diamonds. 

An immediate computational and conceptual advantage of our prescription is the elegant way for alternating between the global AdS and Poincar{\'e} AdS reconstruction formulas of \cite{Hamilton:2006az}, on which we comment in Sec.~\ref{sec:altsmearings}. Rindler reconstruction is not as straightforward in the integral geometric language but we hope to report on it soon.

\subsection{Inverse X-Ray Transform}

\begin{figure}
\centering{\includegraphics[height=0.25\textwidth]{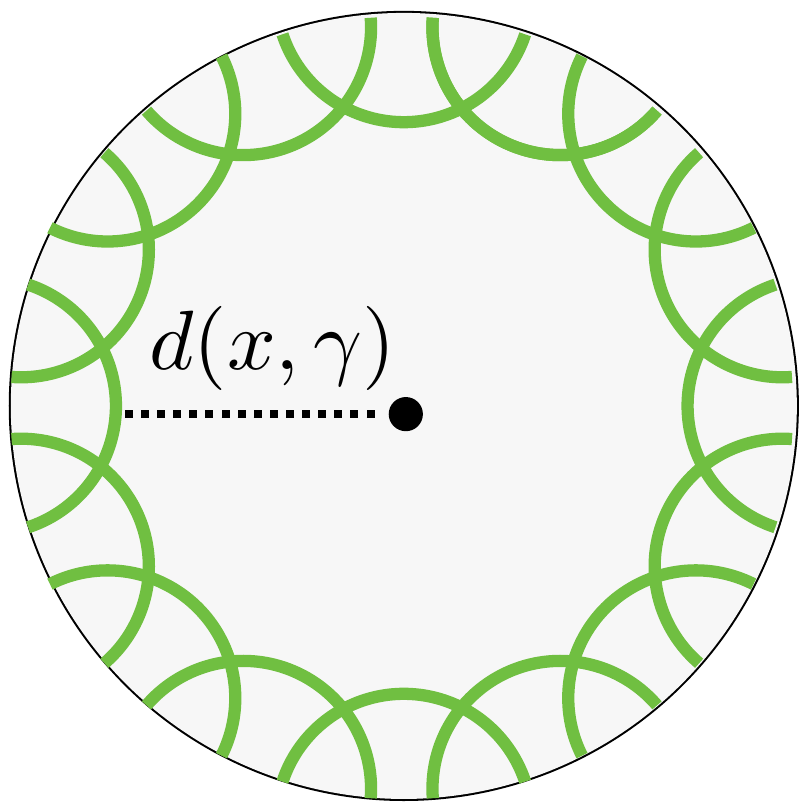}
\caption{When inverting the X-ray transform at a point $x$, we use the average of the transform over geodesics $\gamma$ at a fixed distance $d(x,\gamma)$.  \label{fig:geodesic-to-point-distance}}}
\end{figure}

There are known inversion formulae for Radon transforms over arbitrary-dimension, totally geodesic submanifolds in $\mathbb{H}_d$ \cite{Rubin2002206}. Here, we will only mention the inversion of the X-ray transform in $M=\mathbb{H}_{2}$, since it is the geometry of a time slice of ${\rm AdS}_{3}$, our primary example. 
The inversion formula for the original function $f$ at point $x$ is given by: 
\begin{equation}
f(x)=-\frac{1}{\pi}\int\limits _{0}^{\infty}\frac{dp}{\sinh p}\,\,\frac{d}{dp}\left(\underset{d(x,\gamma)=p}{\text{average}}\,\,Rf\left(\gamma\right)\right).\label{eq:inverseH2}
\end{equation}
This formula asks us to average $Rf\left(\gamma\right)$ over
\emph{all} geodesics at a given proper distance $d(x,\gamma)=p$ from the point $x$ (Fig.~\ref{fig:geodesic-to-point-distance}) 
and then integrate over all distances.
Thus, it requires us to integrate over all geodesics on the hyperbolic slice.

\paragraph{A simple exercise on X-ray transforms}

To get our feet wet with the X-ray transform, let us use (\ref{eq:inverseH2})
to invert the transform for a particularly simple function: $f\left(x\right)=1$.
If we set $f\left(x\right)=1$ in eqn. (\ref{eq:xray-definition}),
we find that $Rf$ is given simply by
\begin{equation}
Rf\left(\gamma\right)=\int_{\gamma}ds=\ell\left(\gamma\right),
\end{equation}
where $\ell\left(\gamma\right)$ denotes the length of the geodesic
$\gamma$. This length is of course infinite, but we can obtain a
cut-off geodesic length instead by setting $f\left(x\right)=\theta\left(\rho_{{\rm uv}}-d\left(x,x_{0}\right)\right)$,
which imposes a radial cutoff $\rho_{{\rm uv}}$ about some center
point $x_{0}$. Doing this for $\rho_{{\rm uv}}\gg1$, we have 
\begin{equation}
\ell_{{\rm cutoff}}\left(\gamma\right)=\log\left(\sin^{2}\alpha\right)+2\rho_{{\rm uv}}
\end{equation}
where $\alpha$ is the opening angle of the geodesic with respect
to the center $x_{0}$. Now the evaluation of (\ref{eq:inverseH2})
for $x=x_{0}$ is straightforward since 
\myeq{\underset{d(x_{0},\gamma)=p}{\text{average}}\,\,Rf\left(\gamma\right)=\ell\left(\alpha\right),}
where we use $p=\sinh^{-1}\cot\alpha$. Then, we have
\begin{equation}
R^{-1} R f (x_0) = \frac{1}{\pi}\int_{0}^{\pi/2}\frac{d\alpha}{\cot\alpha}\frac{d}{d\alpha}\log\left(\sin^{2}\alpha\right)=1
\end{equation}
as expected. Note that allowing the cutoff $\rho_{{\rm uv}}$ to vary
with angle does not change this result, so we can accommodate points
$x\neq x_{0}$ by an equivalent change in cutoff.

\subsection{Global AdS Reconstruction}
\label{sec:globalHKLL}

We now present the holographic construction for a local scalar AdS$_3$ field $\phi$ with mass $m^2=\Delta(\Delta-2)$ in global coordinates. 
We use coordinates $(\rho,\theta,t)$, in which the metric takes the form (we set $L_{\rm{AdS}}=1$):
\myeq{ds^2= -\cosh \rho^2\, dt^2 + d\rho^2 + \sinh \rho^2 \,d\theta^2 \label{metric}\, .}

Recall that the complete kinematic space of AdS$_3$ contains a redundant description of the functions living in the geometry, with different geodesic integrals related by John's equations. 
It is convenient to consider a totally geodesic spacelike slice of AdS$_3$, which has the geometry of two-dimensional hyperbolic space.
The set of X-ray transforms restricted to geodesics on this spatial slice are sufficient to reconstruct functions on the same slice.

We now determine a local operator $\phi(x)$ using the inversion formula for $\mathbb{H}_2$,
\myeq{\phi(x)= -\frac{1}{\pi} \int \limits_{0}^{\infty} \frac{dp}{\sinh{p}}\,\, \frac{d}{dp} \left(\underset{d(x,\gamma)=p}{\text{average}} \,\,\tilde{\phi}(\gamma) \right) \, ,\label{inverseH2}}
where $\tilde \phi(\gamma)$ is the integral of $\phi(x)$ over the geodesic $\gamma$. 
Because our procedure is manifestly invariant under conformal transformations, we need only reconstruct $\phi(\rho,\theta,t)$ at the origin of AdS$_3$ ($\rho=0$) and at time $t=0$. 
The operator at different points can be constructed by appropriate application of bulk isometries.
As mentioned above, the inversion formula identifies a bulk point in a gauge-invariant way: by the distance of all geodesics to the point. 

\begin{figure}
\centering{\includegraphics[height=0.25\textwidth]{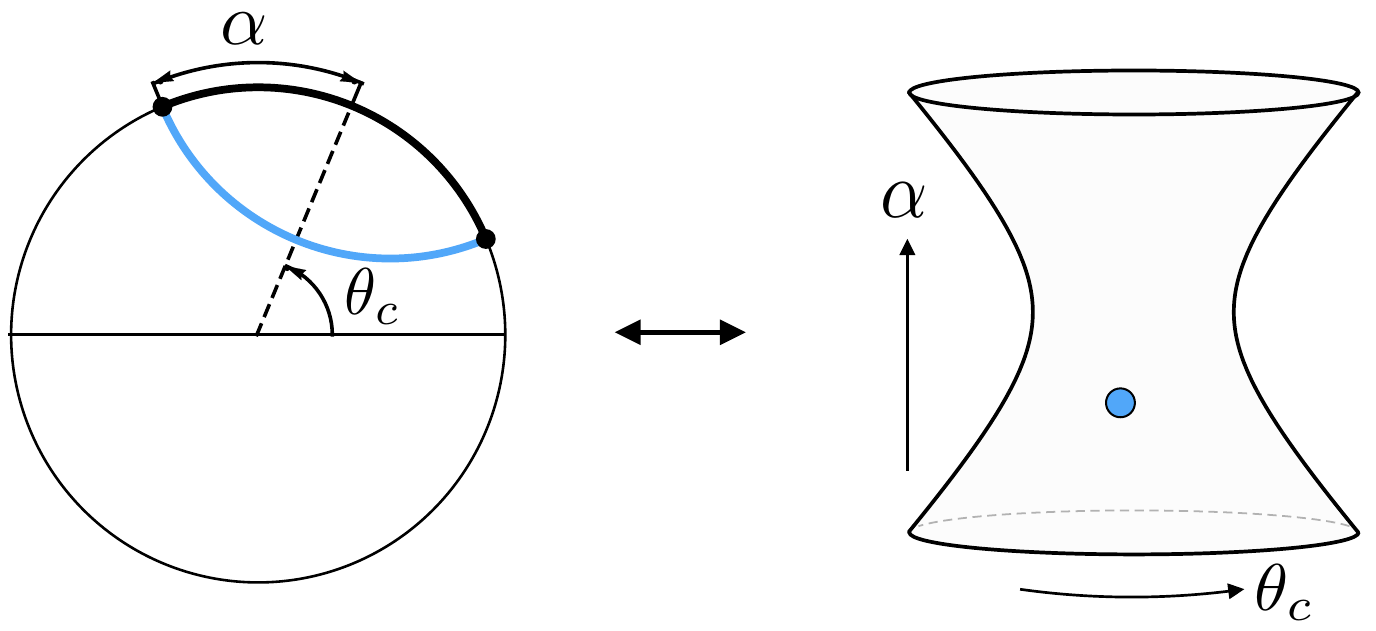}
\caption{We parameterize the kinematic space for $\mathbb{H}_2$ by the opening angle $\alpha$ and midpoint $\theta_c$ of the geodesic.  \label{fig:h2-ks}}}
\end{figure}

\begin{figure}[t]
\centering{\includegraphics[width=1\textwidth]{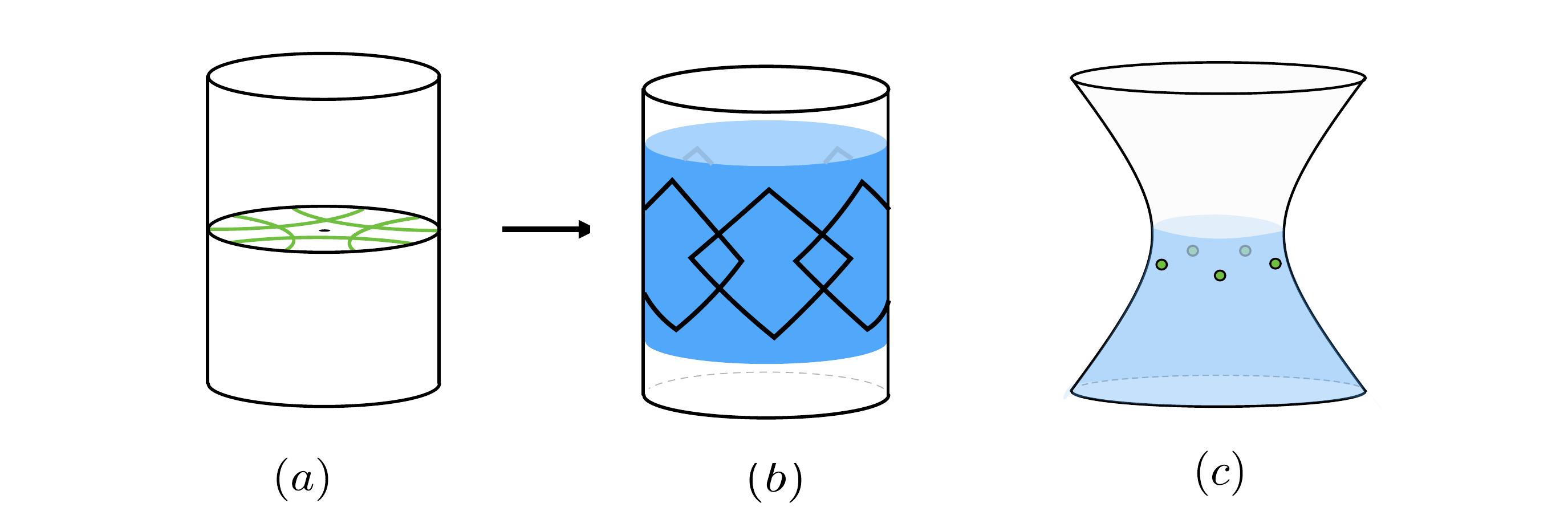}
\caption{(a) A bulk local operator is recovered from an integral over bulk geodesic operators. (b) The boundary representation is a corresponding integral over diamond-smeared operators.  The result is a smeared representation of the bulk local operator that is supported on a time-interval of the cylinder tiled by these diamonds. (c) For each geodesic, we chose the causal diamond that subtends less than half the circle. The domain of integration is the lower-half of the kinematic space for the hyperbolic plane.  A few corresponding examples are shown in each panel. \label{fig:global-smearing}}}
\end{figure}

We will parameterize geodesics in global coordinates by the location of their center $\theta_c$ and their boundary opening angle $\alpha$; see Fig.~\ref{fig:h2-ks}.
Adapted to this coordinate choice, the inversion formula becomes:
\myeq{\phi(\rho=0)= \frac{1}{2\pi^2} \int \limits_0^{2\pi} d\theta_c \int \limits_{0}^{\pi/{2}} d\alpha\,\,\tan \alpha \frac{d}{d\alpha} \tilde{\phi}(\alpha,\theta_c) \, .\label{globalinverse}}
We can now use the CFT representation of the X-ray transform of a local bulk operator found in eq.~\eqref{kinematicdictionary}, $\tilde \phi (\alpha, \theta_c) = c_\Delta^R \mathcal{B}_{\Delta}(\alpha,\theta_c)$ with $c_\Delta^R = \Gamma\left(\frac{\Delta}{2}\right)^2/2\Gamma(\Delta)$, to re-express the bulk local field in terms of boundary operators. 
The OPE block (\ref{eq:smeared}) for the family $h=\bar{h}=\Delta/2$ at $t_1=t_2=0$ can be rewritten in global coordinates as
\myeq{\mathcal{B}_{\Delta}(\alpha,\theta_c) = c_\Delta^{\mathcal{B}} \int_\diamond d\theta dt\,\left(2\frac{\left(\cos t- \cos\left(\theta-\theta_c+\alpha\right)\right)\left(\cos t- \cos\left(\theta-\theta_c-\alpha\right)\right)}{1-\cos\left(2\alpha\right)}\right)^{\frac{\Delta}{2}-1}\OO_{\Delta}(t,\theta) \label{globalblock}}
with $c_\Delta^{\mathcal{B}} = 2\left(\frac{\Gamma(\Delta)}{\Gamma\left(\frac{\Delta}{2}\right)^2}\right)^2$.
Substituting this formula into eq.~\eqref{globalinverse}, we can reverse the order of integration so that we integrate over the geodesic parameters in the inversion formula while leaving the boundary spatial coordinates from the OPE block unintegrated. 
Having done so, eq.~\eqref{globalinverse} takes the form (see also Fig. \ref{fig:global-smearing})
\myeq{\phi(\rho=0)= \int\limits_{-\pi/{2}}^{\pi/{2}} dt\int\limits_0^{2\pi} d\theta\, K_{\Delta}(t)\,\,\OO_{\Delta}(t,\theta) \, ,}
where the smearing function $K_{\Delta}(t)$ is given by the integral expression:
\myeq{K_{\Delta}\left(\tau\right)= \frac{c_\Delta^R c_\Delta^{\mathcal{B}}}{2 \pi^2}\int_{\left|\tau\right|}^{\pi/2}d\alpha\,\tan\alpha\frac{d}{d\alpha}\int_{-\left(\alpha-\left|\tau\right|\right)}^{\alpha-\left|\tau\right|}d\phi\left[2\frac{\left(\cos\tau-\cos\left(\phi+\alpha\right)\right)\left(\cos\tau-\cos\left(\phi-\alpha\right)\right)}{1-\cos\left(2\alpha\right)}\right]^{\Delta/2-1}}
The integral is divergent when evaluated at the upper limit of integration. 
This divergence is of UV nature in the bulk: the set of geodesics with half-width $\alpha=\frac{\pi}{2}$ are precisely the AdS diameters which intersect at the origin $\rho=0$, and they determine the point we are reconstructing. 
We can regulate this divergence by cutting off the integral at $\alpha=\frac{\pi}{2}-\epsilon$ and take the limit $\epsilon\rightarrow 0$ at the end. 
We will see that the bulk operator is insensitive to the regulator.

\begin{figure}[t]
\centering{\includegraphics[width=0.8\textwidth]{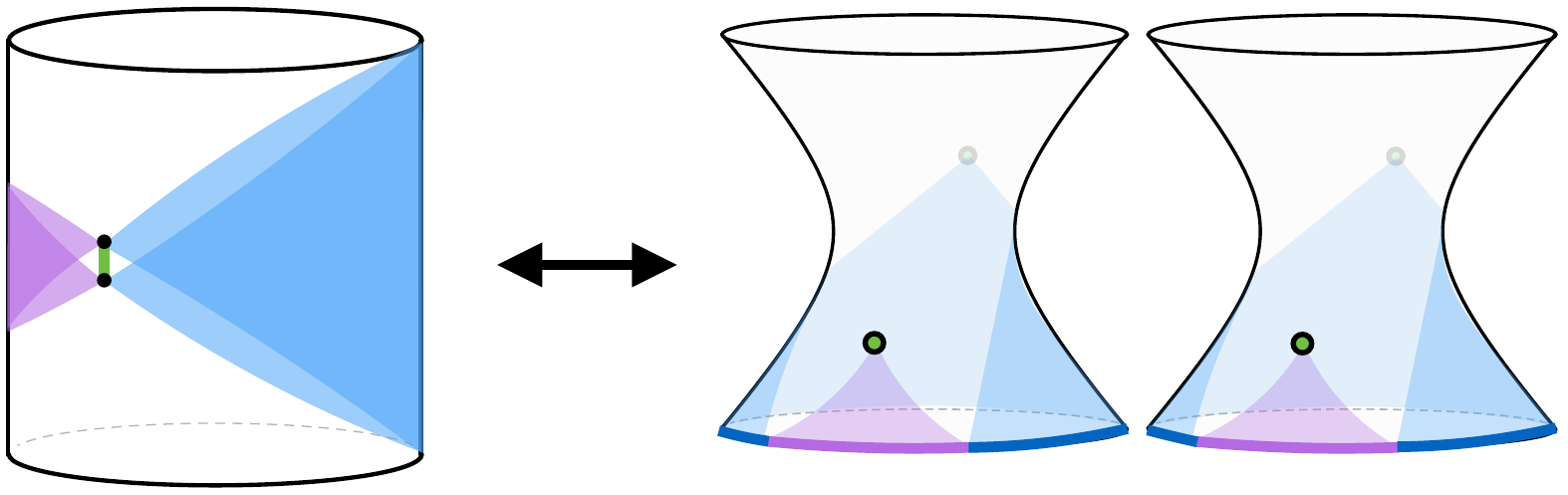}
\caption{In the global $\rm{AdS}_3$ geometry, there are two choices of boundary causal diamond corresponding to a given spacelike geodesic.  These correspond to the two orientations of the geodesic, or equivalently the ordering of the boundary endpoints.  The geodesic operator then has two different representations as a smeared boundary operator.  \label{fig:complementary-diamonds}}}
\end{figure}

The regularized smearing function is computed to be
\myeq{K_{\Delta}(t)= \frac{2^{\Delta-2}\left(\Delta-1\right)}{\pi^2} (\cos t)^{\Delta-2}\left( \log \cos{t} -\log \epsilon -\psi(\Delta-1) - \gamma \right) \, ,}
where $\psi(n)$ is the digamma function. 
The divergent term appears worrisome, but, inserted into the integral with $\mathcal{O}_\Delta(t,\theta)$, the constant terms in the brackets give vanishing contribution as their Fourier expansion has no overlap with the operator. 
They can thus be safely discarded.\footnote{This is exactly analogous to the procedure originally carried out by HKLL to derive smearing functions for bulk operators \cite{Hamilton:2006az}.}
We conclude that the inversion formula determines a boundary smearing function for the bulk operator given by:
\begin{equation}
K_{\Delta}(t)= \frac{2^{\Delta-2}\left(\Delta-1\right)}{\pi^2} (\cos t)^{\Delta-2}\log \cos{t}  \, .
\label{eq:}
\end{equation}
The region of integration is depicted in Fig.~\ref{fig:global-smearing}. This is the same smeared representation of a bulk operator at the center of AdS$_3$ as that found by HKLL \cite{Hamilton:2006az}.

\subsection{Different Smearings}
\label{sec:altsmearings}

\begin{figure}
\centering{\includegraphics[width=1\textwidth]{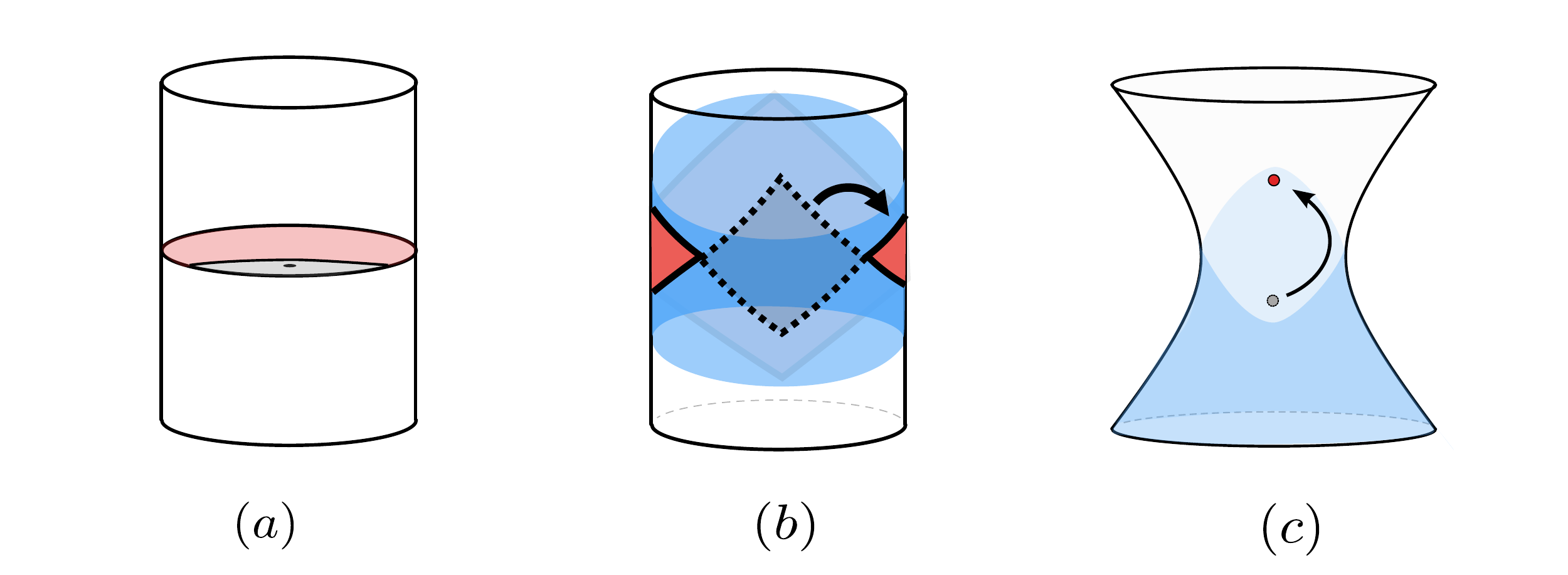}
\caption{(a) We choose the orientation for the geodesic so that the enclosed region does not contain a specified bulk point. (b) The boundary representation is a corresponding integral over the region spacelike-separated from the bulk point. (c) The point identifies half of kinematic space as our domain of integration. The boundary of this region is precisely the `point-curve' of \cite{Czech:2015qta}: the geodesics that intersect at the point. \label{fig:global-smearing2}}}
\end{figure}

There is one subtle puzzle with our derivation: the smearing function we generated had spacelike support from the centre of AdS$_3$ because we integrated over the complete set of geodesics. 
If we were to choose a different point, we would still integrate over the same set of geodesics, and our smearing function would then have spacelike support from the center, not from the chosen point. 
This makes it hard to see how our formula will transform under symmetries to remain the same as that found by HKLL.

The resolution to this puzzle is that there was an implicit choice in the OPE blocks that we used. 
The inversion formula requires an integral over the space of geodesics, without orientation, but our OPE blocks contain an orientation (the choice of one causal diamond or its complement -- see Fig.~\ref{fig:complementary-diamonds}).
We must integrate over all geodesics, but we are free to choose which half of kinematic space we want. 
To obtain the spacelike Green's function, we choose each causal diamond so that the region enclosed by it and the geodesic do not contain the specified bulk point (see Fig.~\ref{fig:global-smearing2}). 
In result, we obtain an integral supported on the boundary region that is spacelike separated from the identified bulk point. 
This is precisely what would have happened had we used conformal transformations to move the point at the center of AdS. 

\begin{figure}
\centering{\includegraphics[width=0.6\textwidth]{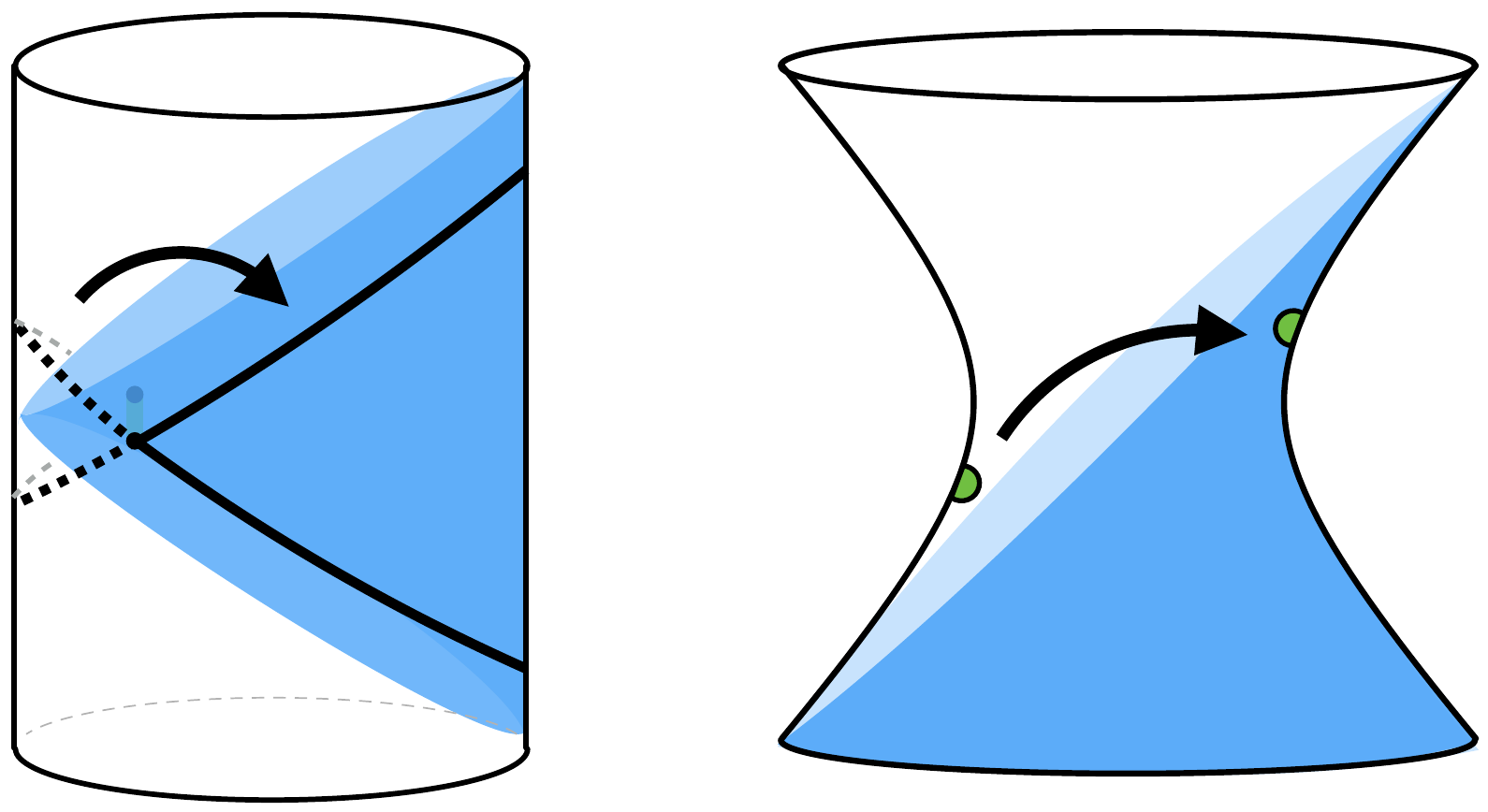}
\caption{If we wish to obtain the Poincar\'{e} representation of a bulk operator, we can make use of the redundancy in the representation of the geodesic operators used to construct it.  To do so, choose for each geodesic the boundary diamond representation that is contained in the desired patch. \label{fig:global-vs-poincare}}}
\end{figure}

A nice feature of our procedure is that the Poincar{\'e} smearing function appears as just another choice of orientation for our OPE blocks. Specifically, the Poincar{\'e} smearing function arises from the choice of OPE block orientations in which none of the causal diamonds contains a fixed boundary point; see Fig.~\ref{fig:global-vs-poincare}.

\paragraph{Other extensions} Our reconstruction of local operators can be extended both to higher dimensions (see Sec.~\ref{higherd}) and to interacting fields. We will sketch how to include interactions in the Discussion, leaving a more complete treatment to future work.

\section{Further Applications}
\label{apps}

\subsection{Vacuum Modular Hamiltonian}
We begin the discussion of applications of our ``kinematic dictionary''
by considering an example of special interest: the OPE block built
out of the CFT stress tensor. We again focus our attention on ${\rm CFT}_{2}$;
the extension to higher dimensions requires the extra machinery presented
in Sec.~\ref{higherd} and we discuss it there.

The stress tensor in two-dimensional CFTs has two independent components of dimension
$\Delta=2$. They are conventionally defined as $T\left(z\right)=-2\pi T_{zz}\left(z\right)$
and similarly for $\bar{T}\left(\bar{z}\right)$, with spin $\ell=2$
and $\ell=-2$, respectively. Recalling the smeared representation
(\ref{eq:smeared}) of OPE blocks we can construct two kinematic fields:
\begin{eqnarray}
\mathcal{B}_{T}\left(x_{1},x_{2}\right) & = & 6\int_{z_{1}}^{z_{2}}dw\,\frac{\left(z_{2}-w\right)\left(w-z_{1}\right)}{z_{2}-z_{1}}\,T\left(w\right)\\
\mathcal{B}_{\bar{T}}\left(x_{1},x_{2}\right) & = & 6\int_{\bar{z}_{1}}^{\bar{z}_{2}}d\bar{w}\,\frac{\left(\bar{z}_{2}-\bar{w}\right)\left(\bar{w}-\bar{z}_{1}\right)}{\bar{z}_{2}-\bar{z}_{1}}\,\bar{T}\left(\bar{w}\right).\nonumber 
\end{eqnarray}
Note that since $T\left(w\right)$ has no dependence on $\bar{w}$,
the $\bar{w}$ integral and its associated normalization factor cancel out
in $\mathcal{B}_{T}$, and similarly for $\mathcal{B}_{\bar{T}}$.

In our conventions, the stress tensor couples to a CFT scalar $\mathcal{O}$
with OPE coefficients $C_{\mathcal{O}\mathcal{O}T}=C_{\mathcal{O}\mathcal{O}\bar{T}}=\frac{\Delta_{\mathcal{O}}}{c}$.
Hence, the stress tensor OPE blocks appear in the $\mathcal{O}\left(x_{1}\right)\mathcal{O}\left(x_{2}\right)$
operator product in the symmetric combination $\mathcal{B}_{T}+\mathcal{B}_{\bar{T}}$.
This sum of blocks can be simplified and brought to a suggestive form.
The energy density can be written as 
\begin{equation}
T_{00}\left(z,\bar{z}\right)=-\frac{1}{2\pi}\left(T\left(z\right)+\bar{T}\left(\bar{z}\right)\right),
\end{equation}
Now, in the simple case where $x_{1}$ and $x_{2}$ lie on the same
time slice, we have 
\begin{equation}
\mathcal{B}_{T}+\mathcal{B}_{\bar{T}}=-12\pi\int_{x_{1}}^{x_{2}}dx\frac{\left(x_{2}-x\right)\left(x-x_{1}\right)}{x_{2}-x_{1}}T_{00}\left(x\right)
\end{equation}
where $T_{00}$ is integrated along the interval that connects the
two points. The result for arbitrary $x_{1},x_{2}$ can be obtained
by applying a boost.

Apart from a normalization mismatch, the stress tensor block is identical
to the \emph{modular Hamiltonian} for the vacuum state \cite{Casini2011}:
\begin{equation}
\mathcal{B}_{T}+\mathcal{B}_{\bar{T}}=-6H_{{\rm mod}}.
\end{equation}
Indeed, this result implies that the modular Hamiltonian appears in
the OPE of any two CFT scalars of equal dimension:
\begin{equation}
\mathcal{O}\left(x_{1}\right)\mathcal{O}\left(x_{2}\right)=\frac{1}{\left|x_{1}-x_{2}\right|^{2\Delta}}\left(1-\frac{6}{c}\Delta_{\mathcal{O}}\,H_{{\rm mod}}+\ldots\right).
\end{equation}
Let us apply this to the twist operators $\sigma_{n}^{\dagger},\sigma_{n}$ of dimension $\Delta=\frac{c}{12}\left(n-\frac{1}{n}\right)$, which
are used in the replica trick computation of the entanglement entropy \cite{Calabrese:2004eu}. Their OPE takes the form
\begin{equation}
\sigma_{n}^{\dagger}\left(x_{1}\right)\sigma_{n}\left(x_{2}\right)=\frac{1}{\left|x_{1}-x_{2}\right|^{\frac{c}{6}\left(n-\frac{1}{n}\right)}}\left(1-\left(n-1\right)H_{{\rm mod}}+\ldots\right),
\end{equation}
where we drop terms of order $\left(n-1\right)^{2}$ and additional
operator contributions. This result was previously noted by \cite{Casini2011}.
Hence, the appearance of the modular Hamiltonian in the OPE is no
accident; the surprise is that it appears so generally.

We will now exploit the fact that the modular Hamiltonian is an OPE
block. This implies that $H_{{\rm mod}}$ is a field on kinematic
space obeying a Klein-Gordon equation (\ref{eq:blockKG}): 
\begin{equation}
\left(2\left(\Box_{{\rm dS_{2}}}+\Box_{\overline{{\rm dS}}_{2}}\right)+4\right)H_{\text{mod}}=0\label{eq:big-entanglement-eom}
\end{equation}
This equation can be combined with the conservation of energy $\partial_{\bar{z}}T(z)=0=\partial_{z}\bar{T}(\bar{z})$
to obtain yet another equation for $H_{\text{mod}}$, which becomes the Klein-Gordon equation on a single de Sitter space: 
\begin{equation}
\left(\Box_{\widetilde{{\rm dS}}_{2}}+2\right)H_{\text{mod}}=0.
\label{eq:small-entanglement-eom}
\end{equation}
This $\widetilde{{\rm dS}}_{2}$ is the ``diagonal'' de Sitter geometry,
which is picked out from the full ${\rm dS_{2}\times\overline{{\rm dS}}_{2}}$
kinematic space by restricting metric (\ref{eq:ksads3}) to $z_{1}=\bar{z}_{1},z_{2}=\bar{z}_{2}$.
In other words, it is the kinematic space for the pairs of points
on a constant time-slice. Given the entanglement first law $\delta S=\left\langle H_{{\rm mod}}\right\rangle $,
eqs.~(\ref{eq:big-entanglement-eom}) and (\ref{eq:small-entanglement-eom}) are entanglement equations of motion.

The same conclusion could be readily reached by observing the form
of the modular Hamiltonian directly. From the perspective of a given
time, e.g. $t=0$, $T_{00}(x,0)$ is a primary of the corresponding
1-dimensional conformal subgroup with weight $\Delta=2$. Moreover,
we can follow again the procedure explained in Sec.~\ref{opeblocks}
and translate the one-dimensional Casimir equation to the Klein-Gordon equation
on the kinematic space of a time-slice. This observation was made
recently in \cite{PhysRevLett.116.061602}. Our key insight is that
this object appears as the OPE contribution of the stress tensor family,
and its apparent propagation in kinematic space constitutes a special
case of a general property of OPE blocks.

Our observation that the stress tensor block equals the vacuum Modular
Hamiltonian will be further developed in \cite{Czech:ee}. It will
enable us to approach the first law of entanglement entropy from a
new perspective and clarify its connection to the bulk linearized
Einstein's equations in the holographic setting, by re-deriving them
in a simpler way. Indeed, we will find that just as scalar equation
of motion intertwine with certain kinematic space equations motion,
so do Einstein's equations intertwine with entanglement equations
of motion.

\subsection{Conformal Blocks} \label{subsec:conformal-blocks}
The identification of OPE blocks with scalar fields in kinematic space provides an elegant geometric description of a fundamental object in the study of CFT correlation functions: the \emph{conformal block}. In view of our kinematic dictionary, this description further allows us to identify the corresponding structure in the holographic dual. Thus, we now turn our attention to CFT four-point functions.

Consider a CFT four-point function $\left\langle \mathcal{O}_{1}\mathcal{O}_{2}\mathcal{O}_{3}\mathcal{O}_{4}\right\rangle$ and make the simplifying assumption that $\Delta_1=\Delta_2=\Delta$ and $\Delta_3= \Delta_4= \Delta'$. Following standard CFT procedure, one can define projection operators $P_{\Delta_k}$ that project any CFT state to states of the $\Delta_k$ irreducible representation. Using the fact that
\myeq{\sum \limits_{k} P_{\Delta_k}=1,}
we decompose the four-point function into ``conformal partial waves'':
\begin{eqnarray}
\left\langle \mathcal{O}_{1}\mathcal{O}_{2}\mathcal{O}_{3}\mathcal{O}_{4}\right\rangle  & = & \sum_{k}C_{12k}C{}_{k34}\mathcal{W}_{k|1234}\left(x_{i}\right). 
\label{eq:4ptfunction}
\end{eqnarray}
Here $C_{ijk}$ are the OPE coefficients and the partial waves are defined by:
\begin{equation}
\mathcal{W}_{k|1234}\left(x_{i}\right)=\frac{1}{C_{12k}C{}_{k34}}\left\langle \mathcal{O}_{1}\mathcal{O}_{2}P_{\Delta_k}\mathcal{O}_{3}\mathcal{O}_{4}\right\rangle  \label{eq:partialwave}
\end{equation}
A given conformal partial wave is the contribution to the four-point function from a specific conformal family of intermediate states. The conformal block $g_{k}\left(u,v\right)$ is defined by convention as 
\begin{equation}
\mathcal{W}_{k|1234}\left(x_{i}\right)=\frac{g_{k|1234}\left(u,v\right)}{x_{12}^{2\Delta}x_{34}^{2\Delta'}}\,, \label{eq:conformalblock}
\end{equation}
where $x_{ij}=x_{i}-x_{j}$.
The variables $u=\frac{x_{12}^{2}x_{34}^{2}}{x_{13}^{2}x_{24}^{2}}$,
$v=\frac{x_{14}^{2}x_{23}^{2}}{x_{13}^{2}x_{24}^{2}}$ are the conformally
invariant cross-ratios. Using the transformation
properties of $\mathcal{W}_{k|1234}$, it is straightforward to verify that $g_{k|1234}$ is indeed conformally invariant.

We can express the partial wave (\ref{eq:partialwave}) in the  language of kinematic space. To do so, recall the definition of OPE blocks as the constituents of the OPE expansion of local operators:
\begin{eqnarray}
\mathcal{O}_{i}\left(x_1\right)\mathcal{O}_{j}\left(x_2\right) & = & \left|x_1-x_2\right|^{-\Delta_{i}-\Delta_{j}}\sum_{k}
C_{ijk}\,\mathcal{B}_{k}^{ij}\left(x_1,x_2\right)
\label{eq:block-definition-2}
\end{eqnarray}  
By expanding the $\mathcal{O}_1\mathcal{O}_2$ and $\mathcal{O}_3\mathcal{O}_4$ products in (\ref{eq:partialwave}) according to (\ref{eq:block-definition-2}) and comparing with the conformal block expression (\ref{eq:conformalblock}), we find that individual conformal blocks become \emph{propagators in kinematic space}:
\begin{equation}
g_{k|1234}\left(u,v\right)=\left\langle 0|\mathcal{B}_{k}\left(x_{1},x_{2}\right)\mathcal{B}_{k}\left(x_{3},x_{4}\right)|0\right\rangle . \label{eq:CBasKSP}
\end{equation}

A subtle issue in our interpretation of conformal blocks as ``kinematic propagators'' arises from the mixed signature of kinematic space, which allows for the construction of various inequivalent propagators. Therefore, one needs to be specific about the choice of propagator consistent with the CFT computation of Lorentzian conformal blocks. The correct answer is obtained by demanding that the propagator have the asymptotic fall-off implied by the OPE block boundary conditions. In effect, the kinematic propagator inherits the singularity structure of conformal blocks. This makes an interesting connection with the signature and discrete symmetries of kinematic space, manifesting its elliptic de Sitter structure \cite{schrodinger, ellipticds2}. We discuss these issues in Appendix~\ref{propagatorchoice}.

\begin{figure}[t!]
\centering{\includegraphics[width=0.8\textwidth]{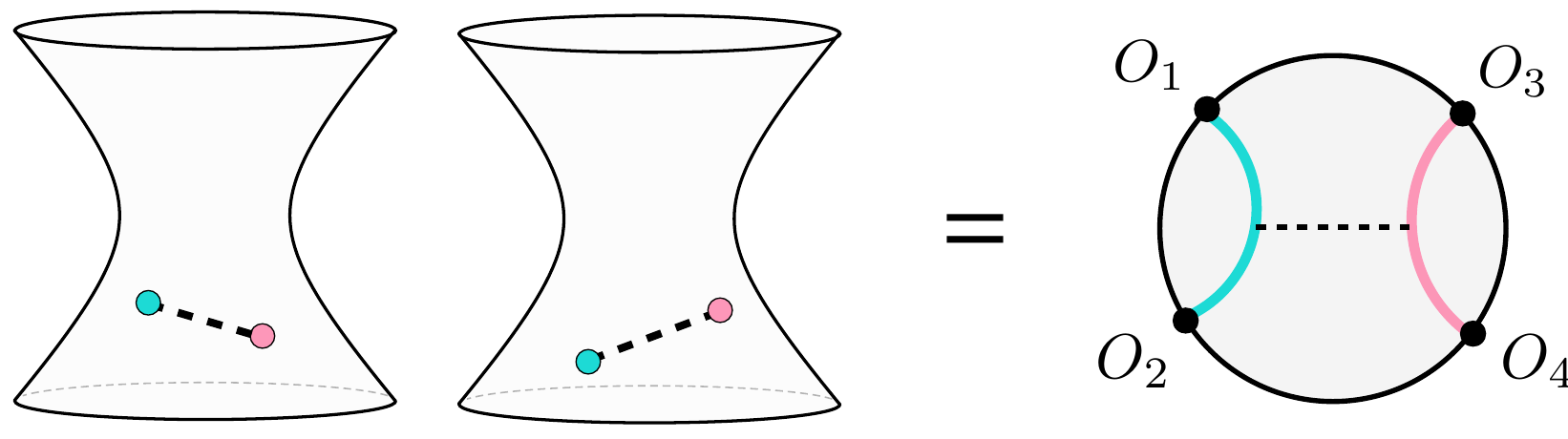}
\caption{Global conformal blocks can be computed equivalently as kinematic space propagators or geodesic Witten diagrams.}
\label{fig:conformal-blocks}}
\end{figure}

\paragraph{Conformal blocks in holography}
The simple representation of conformal blocks as propagators in kinematic space makes it straightforward to identify the corresponding structure in the AdS dual of a holographic CFT. 

Referring to the kinematic dictionary \eqref{kinematicdictionary}, the expression for the conformal block (\ref{eq:CBasKSP}) can be re-written as a correlation function of geodesic integrals of the corresponding bulk field $\phi_{k}$. This yields
\begin{equation}
g_{k|1234}\left(u,v\right)= \frac{1}{c_{\Delta_k}^2} \langle \tilde{\phi}(\gamma_{x_1x_2}) \tilde{\phi}(\gamma_{x_3x_4}) \rangle= \frac{1}{c_{\Delta_k}^2}  \int_{\gamma_{x_1x_2}}ds\int_{\gamma_{x_3x_4}}ds'\,\,G_{bb}(\vec{r}(s), \vec{r'}(s');m_{k})\,, 
\label{eq:geodprop}
\end{equation}
where $G_{bb}(\vec{r}_1, \vec{r}_2;m_{k})$ denotes the bulk-to-bulk AdS propagator for the field dual to the quasi-primary of the $\Delta_k$ conformal family. It is worth noting that in using the dictionary \eqref{kinematicdictionary} to make contact with the bulk field theory, we implicitly assumed that the family $\Delta_k$ is a family of single-trace operators whose AdS duals are local fields at low energies. For multi-trace conformal blocks, there is no corresponding bulk local field and eq. (\ref{eq:geodprop}) is merely a mathematical identity.

The holographic representation of conformal blocks was recently studied in detail in \cite{Hijano:2015,HijanoKrausPerlmutterEtAl2015,Hijano:2015zsa}. The authors recognized that individual conformal partial waves are represented in AdS by ``geodesic Witten diagrams''; see Fig.~\ref{fig:conformal-blocks}. These are similar to the standard Witten diagrams in the bulk, with the difference that the interaction vertices are integrated over geodesics that connect the boundary insertions. As an example, for the simple four-point function (\ref{eq:4ptfunction}) considered in this section, the corresponding partial waves (\ref{eq:partialwave}) are according to \cite{Hijano:2015zsa} equal to:
\myal{\mathcal{W}_{k|1234}(x_i)&= \frac{g_{k|1234}(u,v)}{|x_{12}|^{4\Delta} |x_{34}|^{4\Delta'}} \nonumber\\
	=\frac{1}{c^2_{\Delta_k}}&\int ds \,\int ds'\,\,\,G_{\partial b}(x_1, \vec{r}(s);m_{\Delta}) G_{\partial b}(x_2, \vec{r}(s);m_{\Delta})\times \nonumber\\
	&\times G_{bb}(\vec{r}(s), \vec{r}(s');m_{k}) G_{\partial b}(x_3, \vec{r}(s');m_{\Delta'}) G_{\partial b}(x_4, \vec{r}(s);m_{\Delta'}). \label{eq:geodWitten}}
If we extract the $\Delta_k$ conformal block from this expression, we recover eq.~(\ref{eq:geodprop}), which was derived from an application of our dictionary.


\section{Higher Dimensions}
\label{higherd}

The essential elements of our AdS$_3$/CFT$_2$ arguments remain valid when we go to higher dimensions. There are, however, a number of subtleties, which arise primarily because the conformal group $\rm{SO}(d,2)$ does not factorize like $\rm{SO}(2,2)$ does. The best way to understand these subtleties is to examine the relevant kinematic spaces. 

\subsection{Higher-Dimensional Kinematic Spaces}
\label{sub:higherd-ks}

Right away we come to a fork, because the CFT$_2$ kinematic space can be lifted to higher dimensions in two distinct ways. We saw in Sec.~\ref{ads3kin} that each element of the CFT$_2$ kinematic space labeled one object in the following categories:
\begin{itemize}
\item pairs $(x_1,x_2)$ of space-like separated points in CFT$_2$
\item causal diamonds $\diamond_{12}$
\item pairs of time-like separated points that live on the remaining corners of $\diamond_{12}$
\item bulk geodesics $\gamma_{12}$ in AdS$_3$, which asymptote to $x_1$ and $x_2$ on the boundary
\end{itemize}
Above two boundary dimensions, these concepts depart from one another. A pair of space-like separated points does not define a causal diamond and, consequently, does not select a pair of time-like separated points. To study OPE blocks that arise from products of local operators $\mathcal{O}_i(x_1)$ and  $\mathcal{O}_j(x_2)$, we must therefore distinguish the cases when $x_1$ and $x_2$ are space-like versus time-like separated:

\paragraph{Pairs of space-like separated points $(x_1,x_2)$ in CFT$_d$.} We will denote this space $\mathcal{K}_{g}$. In the presence of a holographic dual, this is also the space of space-like geodesics $\gamma_{12}$ in AdS$_{d+1}$, which end at $x_1$ and $x_2$ on the boundary.

\paragraph{Pairs of time-like separated points in CFT$_d$.}
We will denote this space $\mathcal{K}_{s}$, but for simplicity we do not introduce a new notation for the pair of points $(x_1,x_2)$. The causal cones of $x_1$ and $x_2$ in the CFT intersect on $(d-2)$-dimensional spheres, which bound $(d-1)$-dimensional balls. Thus, we may also think of $\mathcal{K}_{s}$ as the space of causal domains (domains of dependence) of regular $(d-1)$-dimensional balls in CFT$_d$. In a holographic dual, elements of $\mathcal{K}_{s}$ label Ryu-Takayanagi surfaces for boundary balls, denoted $\sigma_{12}$, which are completely homogeneous minimal surfaces in AdS$_{d+1}$.

\begin{figure}
\centering{\includegraphics[width=0.8\textwidth]{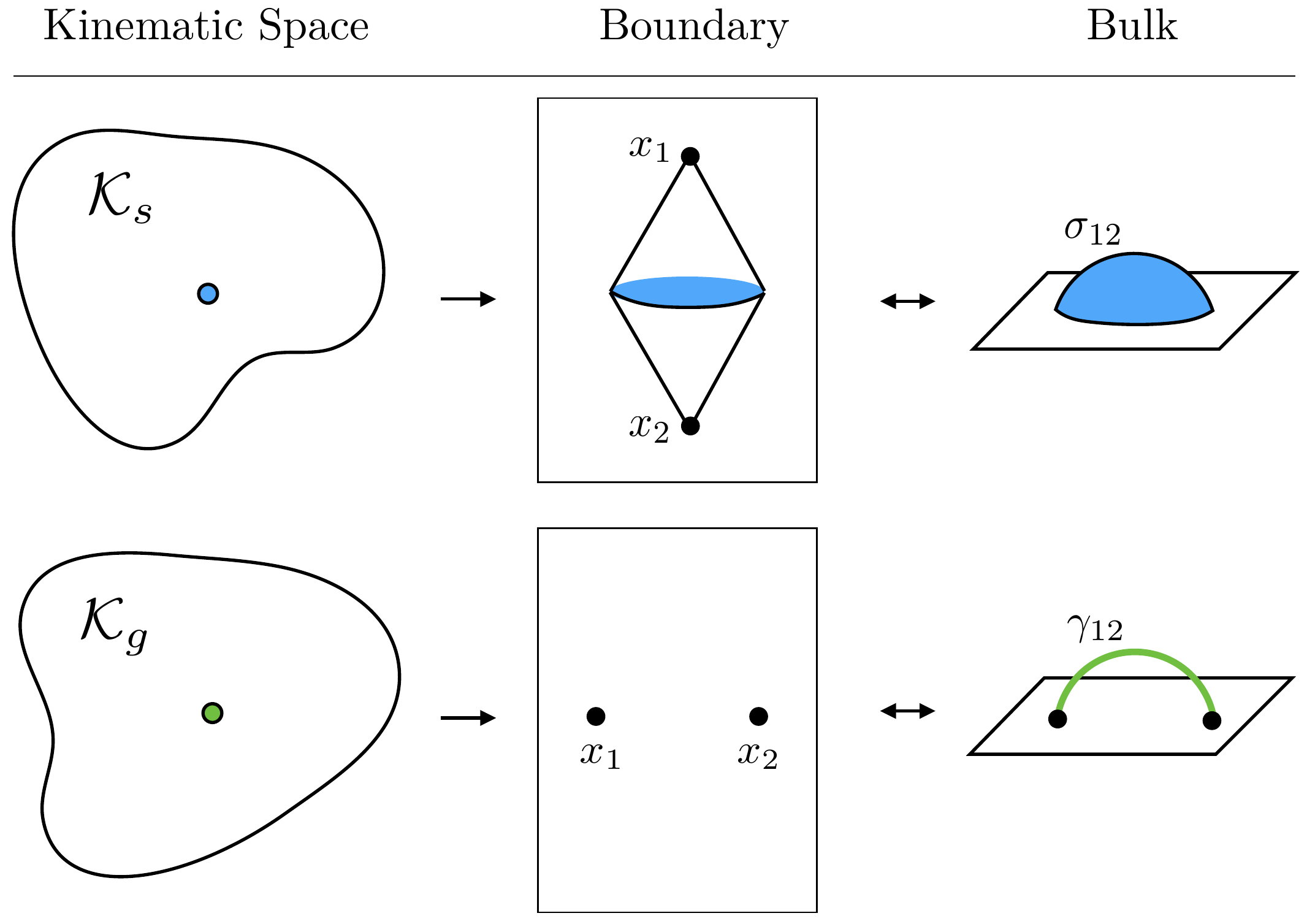}
\caption{The kinematic spaces.\label{fig:twocases}}}
\end{figure}

The two types of kinematic space are illustrated in Fig.~\ref{fig:twocases}. Both inherit a coordinate system from the coordinates of $x_1$ and $x_2$ and are therefore $2d$-dimensional. Because the argument leading to eq.~(\ref{eq:kinmetric}) did not use any tools specific to two-dimensional CFTs, it carries over to the present case. Thus, both $\mathcal{K}_{g}$ and $\mathcal{K}_{s}$ have a unique metric, which is consistent with conformal symmetry:
\begin{equation}
ds^{2}= 
4\left(\eta_{\mu\nu}-2\frac{(x_1-x_2)_{\mu}(x_1-x_2)_{\nu}}{(x_1-x_2)^{2}}\right)
\frac{dx_1^{\mu}dx_2^{\nu}}{\left|x_1-x_2\right|^{2}}.\label{eq:higherdmetric}
\end{equation}
This metric contains $d$ pairs of light-like coordinates corresponding to individual movements of $x_1$ and $x_2$, so its signature is $(d,d)$. 

\subsection{The Radon Transform}

In Sec.~\ref{gops} we considered the X-ray transform on ${\rm AdS}_{3}$,
which takes a function $f$ on ${\rm AdS}_{3}$ to a function
$Rf$ on kinematic space. In higher dimensions more
types of bulk surfaces are available so we can define a wider
variety of transforms.

If we consider the kinematic space $\mathcal{K}_{g}$, we can define
the geodesic X-ray transform $R_g f:\mathcal{K}_{g}\rightarrow\mathbb{R}$
of a function $f:\mathrm{AdS}_{d+1}\rightarrow\mathbb{R}$ as before
by integrating along geodesics:
\begin{equation}
R_g f \left(\gamma\right)=\int_{\gamma}ds\,f\left(x\right).
\end{equation}
Along the same lines, we can consider the kinematic space $\mathcal{K}_{s}$
of codimension-2 minimal surfaces anchored on boundary spheres. The
associated transform is known as the\emph{ Radon transform}, which
we will denote as $R_s f \left(\sigma\right):\mathcal{K}_{s}\rightarrow\mathbb{R}$.
The Radon transform of $f$ is obtained by integrating $f$ over the
bulk surface represented by $\sigma$, weighted by the area element:
\begin{equation}
R_s f \left(\sigma \right)=\int_{\sigma}dA\,f\left(x\right).
\end{equation}
Both the X-ray transform and the Radon transform are known to be invertible
in flat space and hyperbolic space.

\subsection{Kinematic Equations of Motion}

As before, we would like to solve the boundary value problem for a
free scalar field $\phi$ in ${\rm AdS}_{d+1}$, satisfying the Klein-Gordon
equation
\begin{equation}
\left(\square_{{\rm AdS}}-m^{2}\right)\phi\left(x\right)=0.
\end{equation}
 The boundary data is specified by the AdS/CFT dictionary:
\begin{equation}
\phi\left(z\rightarrow0,x\right)\sim z^{\Delta}\mathcal{O}_{\Delta}\left(x\right),\label{eq:highd-dictionary}
\end{equation}
where $\mathcal{O}_{\Delta}$ is the CFT operator corresponding to
the bulk field $\phi$. Since the boundary data is specified on a
codimension-1 surface of ${\rm AdS}_{d+1}$, only a single additional
equation is required to pose a meaningful boundary value problem.
The Klein-Gordon equation fills this role.

However, note that the X-ray and Radon transforms take a function
of $d+1$ variables to a function of $2d$ variables. The boundary
data is still a function of $d$ variables, so now we require $d$
equations to pose the boundary value problem. In both cases, one equation
will come from the Klein-Gordon equation by an intertwining relation.
We will find that the remaining $d-1$ equations take the form of
constraint equations.

\paragraph{Intertwinement}

First, we turn our attention to the intertwining relation. In Sec.~\ref{gops}, we proved that the ${\rm AdS}_{3}$ Laplacian and the
kinematic space Laplacian intertwine under the X-ray transform:
\begin{equation}
\square_{\mathcal{K}} R f= - R \square_{{\rm AdS}_{3}}f.
\end{equation}
Looking back, we can see that the proof also applies without modification
to the X-ray and Radon transforms in higher-dimensional AdS. Indeed,
the fact that ${\rm AdS}_{d+1}$, $\mathcal{K}_{g}$, and $\mathcal{K}_{s}$
are all homogeneous spaces of the group $G={\rm SO}\left(d,2\right)$
implies that the Casimir operator of $G$ is represented by some multiple
of the Laplacian on each (see Appendix \ref{homogeneous}). As a result, the AdS Laplacian
intertwines with the $\mathcal{K}_{g}$ and $\mathcal{K}_{s}$ Laplacians,
respectively, for the X-ray and Radon transforms: 
\begin{eqnarray}
\square_{\mathcal{K}_{g}} R_g f& = & - R_g \square_{{\rm AdS}}f \nonumber \\
\square_{\mathcal{K}_{s}} R_s f & = & - R_s \square_{{\rm AdS}}f.
\end{eqnarray}
This immediately implies that the X-ray or Radon transforms of a free
scalar field propagating in AdS of mass $m$ become  free scalars
on their respective kinematic geometries:
\begin{equation}
\left(\square_{{\rm AdS}}-m^{2}\right)\phi\left(x\right)=0\quad\implies\quad{\left(\square_{\mathcal{K}_{s}}+m^{2}\right)R_s \phi\left(\sigma\right)=0\atop \left(\square_{\mathcal{K}_{g}}+m^{2}\right)R_g \phi\left(\gamma\right)=0}
\end{equation}
This provides the first out of the $d$ necessary equations.

\paragraph{John's equations}

For the remaining $d-1$ equations, we again look for an analog of
John's equations. In Sec.~\ref{gops} we found that the difference
of the two ${\rm dS}_{2}$ Laplacians annihilates the X-ray transform
of any scalar function on ${\rm AdS}_{3}$:
\begin{equation}
\left(\square_{{\rm dS}_{2}}-\square_{\overline{{\rm dS}}_{2}}\right) Rf \left(\gamma\right)=0.\label{eq:ads3-spin-casimir}
\end{equation}
This equation comes from the Casimir operator $S=L_{L}^{2}-L_{R}^{2}$,
which is composed of the Casimir operators of the two factors of ${\rm SO}\left(2,2\right)={\rm SO}\left(2,1\right)_{L}\times{\rm SO}\left(2,1\right)_{R}$. 

Now note that the $d$-dimensional conformal group ${\rm SO}\left(d,2\right)$
has a subgroup ${\rm SO}\left(2,2\right)$ corresponding to the conformal
transformations fixing any ${\rm AdS}_{3}$ slice of ${\rm AdS}_{d+1}$.
We would like to consider such slices that contain two boundary points
$x_1,x_2$ corresponding to an element of kinematic space. Such a slice
can be specified by two boundary vectors $v_{1},v_{2}$ at $x_1$. A
conformal transformation can then be applied such that these vectors
span a plane containing $x_1$ and $x_2$. We can form Casimir operators
$S\left(v_{1},v_{2}\right)$ from the two factors of these ${\rm SO}\left(2,2\right)$
subgroups analogous to above. These operators can be be written as
$S_{\mu\nu}\left(x_1,x_2\right)v_{1}^{\mu}v_{2}^{\nu}$ for some collection
of operators $S_{\mu\nu}\left(x_1,x_2\right)$. Note that, since eq.~(\ref{eq:ads3-spin-casimir})
is antisymmetric under exchange of two coordinates, $S_{\mu\nu}$
must also be antisymmetric. 

From our study of ${\rm AdS}_{3}$, we see that the operator $S_{\mu\nu}\left(x_1,x_2\right)$
annihilates the X-ray transform; in other words, it intertwines with
the zero operator:
\begin{equation}
S_{\mu\nu}\left(x_1,x_2\right) R_g f =0.\label{eq:higherd-john-equation}
\end{equation}
Written as a differential, operator $S_{\mu\nu}$ takes the form
\begin{equation}
S_{\mu\nu}\left(x_1,x_2\right)=I_{\mu}^{\alpha}\left(x_1-x_2\right)\frac{\partial^2}{\partial x_1^{\alpha}\partial x_2^{\nu}}-I_{\nu}^{\alpha}\left(x_1-x_2\right)\frac{\partial^2}{\partial x_1^{\alpha}\partial x_2^{\mu}}\,,\label{eq:higherd-john-operator}
\end{equation}
where $I_{\mu}^{\alpha}\left(x_1-x_2\right)$ is the inversion matrix (\ref{invmatrix}). Eq.~(\ref{eq:higherd-john-operator}) can be checked by noting that it reduces to (\ref{eq:ads3-spin-casimir}) for the case of $d=2$.

Eq.~(\ref{eq:higherd-john-equation}) is what we call the AdS John's equation.
It bears a striking resemblance to John's equations in flat space
\cite{john1938}, which completely characterize the image of the scalar X-ray transform. It is natural to conjecture
that the same holds true in AdS.

Since we have shown that the operator corresponding to $S_{\mu\nu}\left(x_1,x_2\right)$
is represented by the zero operator on scalar AdS functions, it follows
from the intertwining relation (\ref{generator-intertwinement}) that $S_{\mu\nu}\left(x_1,x_2\right)$
also annihilates the Radon transform:
\begin{equation}
S_{\mu\nu}\left(x_1,x_2\right) R_s f =0.
\end{equation}
Hence, John's equations apply both to the X-ray and the Radon transforms.

Note that, since $S_{\mu\nu}$ is an antisymmetric $d\times d$ matrix,
John's equation (\ref{eq:higherd-john-equation}) actually consists
of $d\left(d-1\right)/2$ separate equations. We expect that,
as in flat space, only $d-1$ of these equations are independent, and that
they completely characterize the range of the X-ray and Radon transforms.

\paragraph{Boundary conditions}

To pose the boundary value problem for the X-ray and Radon transforms,
we must again specify boundary conditions. These can be obtained
as before by considering a geodesic or surface near the boundary of
kinematic space and applying the AdS/CFT dictionary (\ref{eq:highd-dictionary}).

The boundary conditions for the X-ray transform are as before:
\begin{equation}
R_g \phi\left(\gamma_{12} \right) \underset{x_2 \rightarrow x_1}{\rightarrow} \left(\int_{\gamma_{12}}ds\,z^{\Delta}\right)\mathcal{O}\left(x_1 \right)= c_\Delta^g \left|x_1-x_2\right|^{\Delta}\mathcal{O}\left(x_1\right);  \qquad c_\Delta^g = \frac{\Gamma\left(\frac{\Delta}{2}\right)^2}{2\Gamma\left(\Delta\right)}
\end{equation}
For the Radon transform, the boundary
conditions are
\begin{equation}
R_s \phi\left(\sigma_{12}\right) \underset{x_2 \rightarrow x_1}{\rightarrow} \left(\int_{\sigma_{12}}dA\,z^{\Delta}\right)\mathcal{O}\left(x_1\right)= c_\Delta^s \left|x_1-x_2\right|^{\Delta} \mathcal{O}\left(x_1\right);  \quad c_\Delta^s = \frac{\pi^{\frac{d-1}{2}}\Gamma\left(\frac{\Delta-d+2}{2}\right)}{2^\Delta \Gamma\left(\frac{\Delta+1}{2}\right)}.
\end{equation}
These boundary conditions, together with the Klein-Gordon equation
and John's equations, determine the X-ray and Radon transforms.

\subsection{Bilocals and Surface Operators}

We will now relate these geodesic and surface operators to higher dimensional OPE blocks.  Recalling Sec.~\ref{opeblocks}, the discussion of the OPE expansion in ${\rm CFT}_{2}$ applies equally well in
higher dimensions. 
Scalar OPE blocks in higher-dimensions also obey the Casimir equation and John's equations, with the boundary conditions:
\begin{equation}
\mathcal{B}_{k}\left(x_1,x_2\right)\underset{x_2\rightarrow x_1}{\rightarrow}\left|x_1-x_2\right|^{\Delta_{k}}\mathcal{O}_{k}\left(x_1\right) \, .
\end{equation}
By matching this data to our bulk calculations, we conclude that
the higher-dimensional OPE blocks are the CFT representations of geodesic
and surface operators: 
\begin{equation}
\mathcal{B}_{k}\left(x_1,x_2\right)=\begin{cases}
\frac{1}{c_\Delta^g} \int_{\gamma_{12}} ds\phi_{k} & \left(x_1,x_2\right)\text{ spacelike}\\
\frac{1}{c_\Delta^s}  \int_{\sigma_{12}} dA\,\phi_{k} & \left(x_1,x_2\right)\text{ timelike}
\end{cases} \label{eq:ope-block-dictionary-higher-d}
\end{equation}
In the timelike-separated case, it may be more useful to think of these blocks as being the contribution of a conformal family, not to local operators inserted at timelike-separated points $(x_1,x_2)$, but to a surface operator $\Sigma(x_1,x_2)$ localized on the intersection
of the light-cones of the points $x_1,x_2$. Let us elaborate on this now.

\paragraph{Surface operator OPE}

Just as a bilocal operator can be expanded in terms of a local operator basis using the state-operator correspondence, so too can a surface operator. 
The surface operator expansion is particularly relevant, as spherical twist operators are used to calculate Renyi entropies in higher dimensions \cite{Hung:2014npa}. 

We will consider a scalar surface operator $\Sigma$$\left(x_1,x_2\right)$ localized
on a boundary $d-2$-sphere defined by the two points.   
We can expand such a surface operator in terms of surface OPE blocks:
\begin{equation}
\Sigma\left(x_1,x_2\right)=\left\langle \Sigma\left(x_1,x_2\right)\right\rangle 
\sum_{i}c_{i}\,\mathcal{B}_{i}^{s}\left(x_1,x_2\right).
\end{equation}
Here the $c_{i}$ are constant coefficients that depend on the choice
of operator $\Sigma$. The surface blocks $\mathcal{B}_{i}^{s}$ contain contributions from an entire conformal family. The overall prefactor $\left\langle \Sigma\left(x_1,x_2\right)\right\rangle $
is the vacuum expectation value of the surface operator, which is
assumed to be nonzero. 

\begin{figure}
\centering{\includegraphics[width=0.9\textwidth]{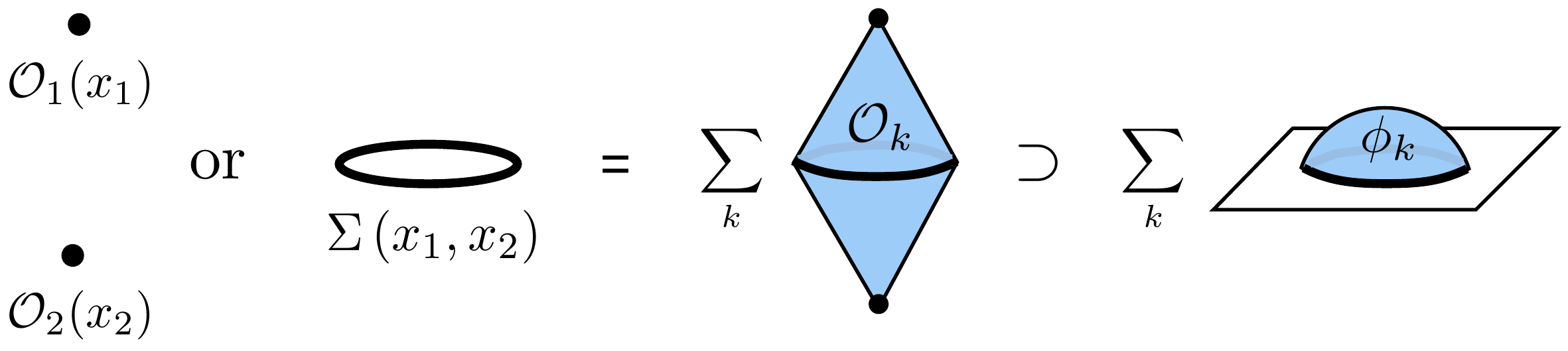}
\caption{OPE blocks and surface operators are equal, and both have a CFT representation as an operator smeared over a causal domain.  OPE blocks appear in the expansion of a bilocal of timelike-separated operators, as well as in the expansion of surface operators.}
\label{fig:OPE-surface}}
\end{figure}

Because the transformation of $\Sigma(x_1,x_2)$ has been completely absorbed into the prefactor, the surface OPE blocks transform as kinematic
space scalars under global conformal transformations:
\begin{equation}
\mathcal{B}_{i}^{s}\left(x_1,x_2\right)\rightarrow\mathcal{B}_{i}^{s}\left(x_1^{\prime},x_2^{\prime}\right).
\end{equation}
Then the logic of Sec.~\ref{opeblocks} tells us that the surface
OPE blocks $\mathcal{B}_{i}^{s}$ obey the same Casimir and John's
equations as the bilocal OPE blocks $\mathcal{B}_{i}$. By normalizing
the coefficients $c_{i}$ appropriately, we can match the boundary
conditions of the surface and bilocal OPE blocks. They must then be
related by analytic continuation:
\begin{equation}
\mathcal{B}_{i}^{s}\left(x_1,x_2\right) \rightarrow \mathcal{B}_{i}\left(x_1,x_2\right).
\end{equation}
We have thus learned that codimension-2 CFT surface operators
can be expanded in terms of the same OPE blocks as timelike-separated
bilocals; see Fig.~\ref{fig:OPE-surface}.

\subsection{Smearing Representations}

Next, we would like to solve the boundary value problem for the higher-dimension
OPE blocks and X-ray/Radon transforms to find a smearing representation
analogous to eq.~\eqref{eq:smeared}.

\paragraph{Causal structure}

First, we must study the causal structure of the higher-dimensional
kinematic spaces $\mathcal{K}_{g}$ and $\mathcal{K}_{s}$. For the
case of ${\rm AdS}_{3}$, the causal structure allowed us to formulate
the Cauchy problem in a more standard way, since the causal past of
kinematic space was a spacelike Cauchy surface on which the boundary
data was specified.

It turns out that $\mathcal{K}_{g}$ and $\mathcal{K}_{s}$ possess
very different causal structures. To see this, it is useful to write
the metric (\ref{eq:higherdmetric}) in ``center of mass'' coordinates $\chi^{\mu}=\frac{x_1^{\mu}+x_2^{\mu}}{2}$
and $\ell^{\mu}=\frac{x_1^{\mu}-x_2^{\mu}}{2}$:
\begin{equation}
ds^{2}=\frac{I_{\mu\nu}\left(\ell\right)}{\left|\ell\right|^{2}}\left(d\chi^{\mu}d\chi^{\nu}-d\ell^{\mu}d\ell^{\nu}\right).
\end{equation}

For $\mathcal{K}_{s}$, $\ell$ is a timelike vector and $I_{\mu\nu}\left(2\ell\right)$
has all positive eigenvalues. Hence, for fixed $\ell$, changes in
$\chi$ are all spacelike. In particular, the $\ell=0$ surface on
which the boundary data is specified is a spacelike surface of dimension
$d$; even the timelike direction on the boundary of
AdS is seen as spacelike in kinematic space. There is also a causal
structure in $\mathcal{K}_{s}$ coming from the containment relation
of boundary causal diamonds and the $\ell=0$ surface sits at the
asymptotic past in this structure. Hence, the Radon transform
converts a non-standard Cauchy problem into a more usual problem like that in ${\rm AdS}_{3}$.

For $\mathcal{K}_{g}$, the structure is not as clear. Here $\ell$
is a spacelike vector so $I_{\mu\nu}\left(2\ell^{\mu}\right)$ has
$d-2$ positive and $2$ negative eigenvalues. This means that the
boundary data is not specified on a spacelike surface and the Cauchy
problem for the X-ray transform is not of a standard type for $d>2$.
The solution to the Cauchy problem will therefore be highly non-unique.

We should not be surprised by this situation, though; it is analogous
to the non-uniqueness of the smearing representation of a bulk local
operator. Just as a point in AdS is contained in many different Rindler
wedges, a geodesic in ${\rm AdS}_{d+1}$ for $d>2$ is contained in
many different wedges and it may have a different representation
in each. A surface in $\mathcal{K}_{s}$, however, forms the boundary
of exactly two Rindler wedges, so there are only two representations
of the surface operator.

\paragraph{Solution to the Cauchy problem}

Now that we have understood the causal structure, we will proceed
to solve the Cauchy problem for the OPE blocks and Radon transform,
yielding a smearing representation of each. Since the OPE block and
Radon transform obey the same equations with the same boundary conditions,
it is actually simplest to find the solution using the shadow operator
formalism of Appendix \ref{app:shadows}. The result is
\begin{equation}
\mathcal{B}_{k}\left(x_1,x_2\right)=n_{k}\int_{\diamond}d^{d}z\left(\frac{\left|y-z\right|\left|x-z\right|}{\left|x_1-x_2\right|}\right)^{\Delta-d}\mathcal{O}_{k}\left(z\right) \; ,\label{eq:highd-ope-block}
\end{equation}
where $n_{k}$ is a normalization factor that depends on $\Delta$
and $d$ and the region of integration is the causal domain
selected by $x_1$ and $x_2$.

We might attempt to extend the result (\ref{eq:highd-ope-block}) to
the X-ray transform or spacelike OPE block. The integrand is still
fixed by conformal invariance and the resultant object still obeys
both the Casimir and John's equations. However, it is no longer
clear how to impose the proper boundary conditions. As $x_1$ approaches
$x_2$, we require that $\mathcal{B}_{k}\left(x_1,x_2\right)$ approach
a local operator. Since the causal diamond approaches a point for
$x_2\rightarrow x_1$, this is easy to do for timelike separated $x_1,x_2$.
However, there is no such small region associated with two spacelike
separated points for $d>2$. Hence, we fail to impose the desired
boundary conditions.\footnote{Note, however, that this affliction does not ruin the equality of
the OPE block and the X-ray transform.}

A particular case of the surface OPE block is the modular Hamiltonian, as was shown for $\rm{CFT}_2$ in Sec. \ref{apps}. Indeed, it can be checked that the OPE block contributions from $T_{\mu\nu}$ and its descendants can be rewritten as precisely the modular Hamiltonian.  The method for computing such tensor OPE blocks is described in Appendix~\ref{app:shadows}.

\subsection{Conformal Blocks and Surface Witten Diagrams}

\begin{figure}
\centering{\includegraphics[width=0.4\textwidth]{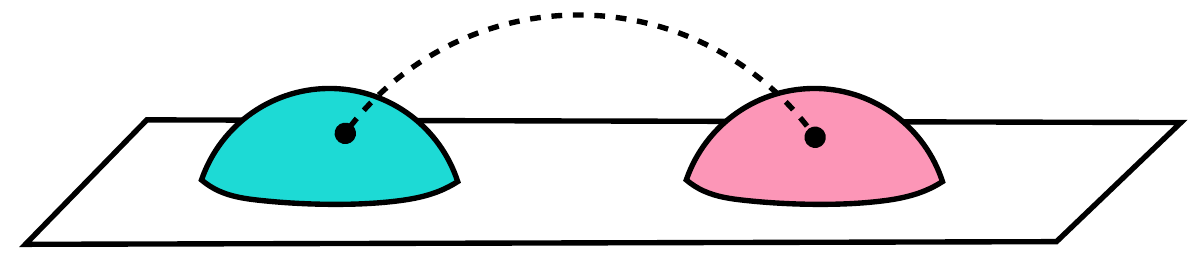}
\hspace{0.05\textwidth}
\includegraphics[width=0.4\textwidth]{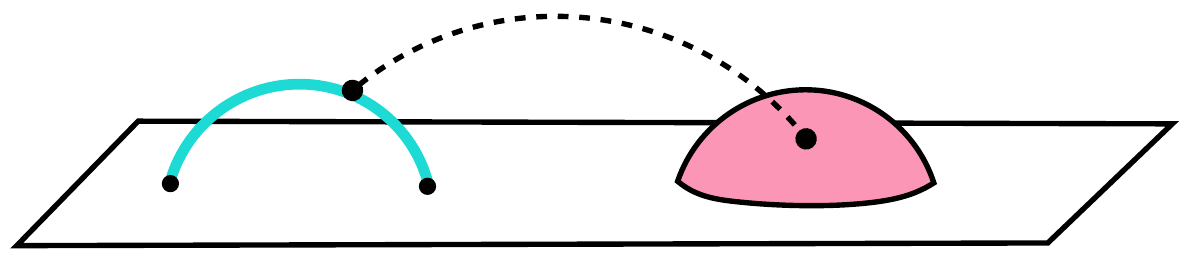}
\caption{Surface Witten diagrams, and mixed surface-geodesic Witten diagrams.\label{fig:new-witten-diagrams}}}
\end{figure}

The equality of OPE blocks and the X-Ray/Radon transforms gives us
an AdS method for computing conformal blocks in higher dimensions.
Indeed, it is immediately clear from the definition of the conformal
block and OPE block that, for $\Delta_{1}=\Delta_{2}$, $\Delta_{3}=\Delta_{4}$,
the conformal block is given by 
\begin{equation}
g_{k}\left(u,v\right)= \left\langle \mathcal{B}_{k}\left(x_{1},x_{2}\right)\mathcal{B}_{k}\left(x_{3},x_{4}\right)\right\rangle .
\end{equation}
Then, just as in Sec. \ref{subsec:conformal-blocks}, the OPE block dictionary \eqref{eq:ope-block-dictionary-higher-d} shows that the higher-dimensional conformal block for spacelike-separated points can be computed by a geodesic Witten diagram \eqref{eq:geodprop} \cite{HijanoKrausPerlmutterEtAl2015,Hijano:2015zsa}.

For the case of timelike-separated endpoints, a new structure emerges.
If we take $x_{1},x_{2}$ and $x_{3},x_{4}$ to be pairwise timelike-separated, the conformal block becomes a \emph{surface} Witten diagram:
\begin{equation}
g_{k}\left(u,v\right)= \frac{1}{\left(c_{\Delta}^s\right)^2}  \int_{\sigma_{12}}d^{d-2}z\int_{\sigma_{34}}d^{d-2}z^{\prime}\,G_{bb}\left(z,z^{\prime};\,m_{k}\right).
\end{equation}
The endpoints of the bulk-to-bulk propagator are now integrated
over a surface rather than a geodesic. Since the OPE block appears
in the expansion of a surface operator, this type of conformal block
also computes a contribution to the correlation function of two surface
operators. If we take $x_{1},x_{2}$ spacelike separated and $x_{3},x_{4}$
timelike separated, the analogous result is a mixed
geodesic/surface Witten diagram. These options are illustrated in Fig.~\ref{fig:new-witten-diagrams}.

Because surface operators can be expanded in terms of the same OPE
blocks as CFT bilocals, a correlation function of surface operators
can be expanded in terms of the same familiar conformal blocks: 
\begin{equation}
\left\langle \Sigma_{i}\left(x_{1},x_{2}\right)\Sigma_{j}\left(x_{3},x_{4}\right)\right\rangle =\left\langle \Sigma_{i}\left(x_{1},x_{2}\right)\right\rangle \left\langle \Sigma_{j}\left(x_{3},x_{4}\right)\right\rangle \sum_{k}c_{ik}c_{jk}g_{k}\left(u,v\right).
\end{equation}
In this expression, $g_{k}\left(u,v\right)$ are the usual conformal blocks while $c_{ik}$ and $c_{jk}$ are the coefficients of the surface operator
expansions.\footnote{A similar result appeared in \cite{Gadde:2016fbj}, which also studied defect operators of general codimension.  This suggests a generalization of kinematic space to include bulk surfaces of arbitrary codimension.} For the case where $\Sigma_{i}$ and $\Sigma_{j}$
are spherical twist operators, this provides a conformal block expansion
for the two-ball Renyi and entanglement entropies. Related formulations
have been explored in \cite{Calabrese:2010he,Cardy:2013nua,Headrick:2010zt}.
In particular, this expansion takes the form of a surface Witten diagram
described above, as was suggested in \cite{Faulkner:2013ana}.

\section{Discussion}

Like any good formulation of a theory of gravity, a holographic CFT describes the bulk via diffeomorphism invariant, and thus inherently non-local, variables. Our goal was to initiate a systematic construction of such variables on both sides of the duality. At the same time, we developed the tools for recovering the familiar, local degrees of freedom.

We adopted a bottom-up approach to this problem. Starting from first principles and working at leading order in $1/N$, we identified an operator correspondence. The relevant CFT object was the OPE block: the non-local operator appearing as the contribution of a single conformal family to the OPE of local operators (or codimension-2 surface operators). OPE blocks were shown to be dual to integrals of the corresponding bulk fields over geodesics (or minimal surfaces), which are also known as Radon transforms. We view this as a stepping stone toward an operator variant of the Ryu-Takayanagi relation.

Our holographic variables, despite their non-locality, admit an elegant geometric description as scalar fields propagating in the auxiliary geometry of kinematic space, defined as the space of CFT pairs of points or, equivalently, the space of bulk geodesics. This structure, which follows from conformal symmetry, underlies the proof of our correspondence. Its implications, however, are more consequential. Recall that kinematic space and integral geometry emerged as essential tools for reconstructing the bulk geometry from entanglement entropies \cite{diffentropy, Czech:2015qta}. With our operator correspondence at hand, we were able to use closely related machinery to assemble local bulk fields from geodesic probes. The conceptual advantage of our approach over the traditional HKLL construction \cite{Banks:1998dd,Balasubramanian:1998de,Bena:1999jv,Hamilton:2005ju,Hamilton:2006az,Heemskerk:2012np,Heemskerk:2012mn} lies in its invariance under diffeomorphisms: both the smearing function and the bulk point are defined with sole reference to the boundary.\footnote{It would also be interesting to compare our construction to Refs.~\cite{Verlinde:2015qfa,Miyaji:2015fia,Nakayama:2015mva}, which associate bulk operators to cross-cap or boundary operators in the CFT.}

The best illustration of the power of the correspondence between OPE blocks and geodesic operators is the range of results, which follow from it with almost no technical effort. In this paper we discussed 
the modular Hamiltonian and its behavior in kinematic space \cite{PhysRevLett.116.061602} as well as the holographic representation of conformal blocks in terms of geodesic Witten diagrams \cite{HijanoKrausPerlmutterEtAl2015,Hijano:2015zsa}. Our upcoming work will explain how the identification of the modular Hamiltonian with an OPE block implies linearized Einstein's equations. We believe that even this set of examples understates the importance of geodesic operators.

Let us sketch some speculative possibilities:


\subsection{Interacting Dictionary}
\label{dis:interactions}
For the purposes of this paper, we limited ourselves to the discussion of free fields in the bulk. The CFT dual to such a theory is trivial because all OPEs are fixed by the factorization of correlation functions and crossing symmetry. Nevertheless, this is the behavior of any CFT with a large $N$ expansion at leading order in $1/N$. At sub-leading orders, the kinematic dictionary (\ref{kinematicdictionary}) admits corrections, which come from bulk perturbative interactions. For a cubic vertex, for example, they read:
\begin{equation}
\tilde\phi(\gamma_{12}) = c_\Delta \mathcal{B}_{\OO_\Delta}(x_1,x_2) + \frac{1}{N} \sum_{\lbrace i,j \rbrace , n } a_n^{ij} \mathcal{B}_{\[\mathcal{O}_i\mathcal{O}_j\]_n}(x_1,x_2)+ O(1/N^2) \, ,
\label{eq:GeoInterOp}
\end{equation}
where 
$[\mathcal{O}_i\mathcal{O}_j]_n$ 
are double-trace primary operators with $2n$ derivatives, constructed from the single-traces $\OO_i$, $\OO_j$ to which $\OO_\Delta$ couples. The $1/N$ corrections appear in the form of double-trace OPE blocks, a fact that renders the computation of the coefficients $a_n^{ij}$ extremely efficient. This corrected operator correspondence can again be combined with the inversion formula for the Radon transform to obtain a smeared CFT representation for interacting bulk fields. We have confirmed that this result is in agreement with \cite{Kabat:2015swa}. An approach to computing $1/N$ corrections to local operators by exploiting the OPE appeared, while this paper was being written, in \cite{Kabat:2016zzr}.

Conceptually, the dictionary in the presence of bulk interactions is especially intriguing. It organizes single-trace and multi-trace OPE blocks in a larger structure, the bulk geodesic operator. This organization of blocks is curiously reminiscent of Virasoro blocks in two-dimensional CFTs, where the multi-trace contributions are formed from the stress tensor. In the latter, the relative coefficients are fixed by local conformal symmetry whereas for the geodesic operator they are associated with the existence of a local perturbative holographic dual at low energies. This suggests a novel organization of the OPE of holographic CFTs in terms of integrals of bulk operators along geodesics. The interacting dictionary will be the subject of an upcoming publication.

\subsection{Beyond the Vacuum}

Though we relied heavily on conformal symmetry throughout this work,
the machinery of integral geometry suggests generalizations to states
beyond the vacuum. For example, kinematic space has proved useful for reconstructing general bulk metrics from boundary data \cite{Czech:2015qta}.

\begin{figure}
\centering{\includegraphics[width=0.6\textwidth]{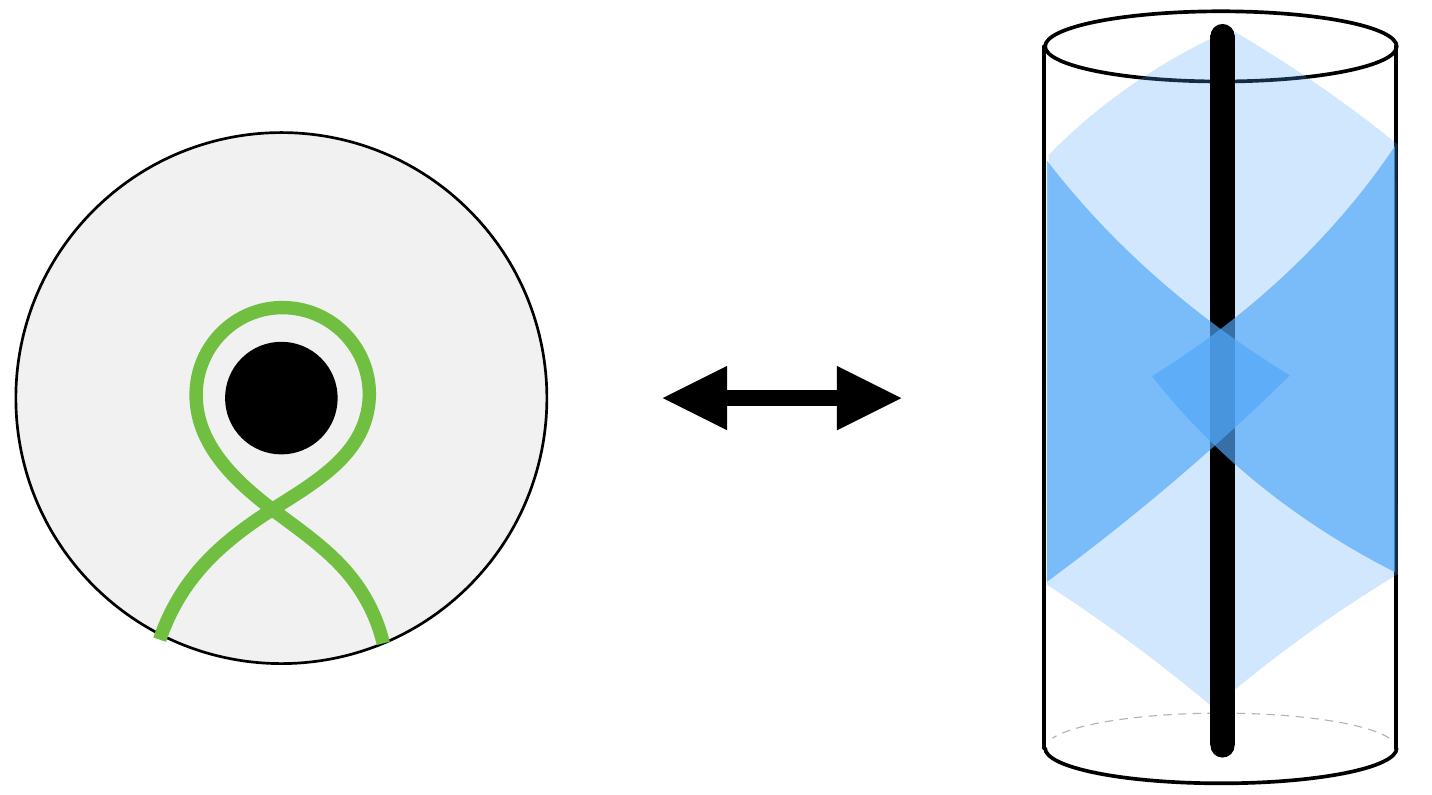}
\caption{Geodesic operators on excited $\rm{CFT}_2$ states can be obtained by taking a quotient both of $\rm{AdS}_3$ and of kinematic space.  Winding (entwinement-type) geodesics \cite{entwinement} correspond to boundary operators smeared over wrapping diamonds.}
\label{fig:winding-geodesics}}
\end{figure}

A particularly simple case arises in ${\rm AdS}_{3}/{\rm CFT}_{2}$,
where a wide class of geometries can be obtained by applying large
bulk diffeomorphisms. Classic examples include the conical defect and
the BTZ geometry, which are simple quotients of ${\rm AdS}_{3}$ \cite{Banados:1992gq}.
Not surprisingly, the corresponding kinematic spaces are also quotients
of the ${\rm AdS}_{3}$ kinematic space, a fact that was exploited in \cite{Czech:2015kbp}.
We can then solve for the geodesic operators just as in the vacuum.
Interestingly, winding (i.e.~entwinement-type \cite{entwinement}) geodesic operators are represented in the CFT
by operators smeared over winding diamonds (see Figure \ref{fig:winding-geodesics}). Moreover, by utilizing the tools of \cite{Fitzpatrick:2015zha}, we are able to prove the result
of \cite{HijanoKrausPerlmutterEtAl2015}: that the semiclassical heavy-light
Virasoro block is computed by a geodesic Witten diagram in a conical defect or BTZ geometry. This direction will be explored in an upcoming paper.

More ambitiously, kinematic space as the space of bulk geodesics is also meaningful in a large class of asymptotically AdS geometries in higher dimensions. It has a notion of volume defined by the Crofton form, which in turn can be interpreted as a ``density of geodesics'' \cite{Helgason1999,9780511617331,Huang:2015aa}:
\myeq{\omega= \frac{\partial^2 S(x_1,x_2)}{\partial x_1^{\mu} \partial x_2^{\nu}}\, dx_1^{\mu}\wedge dx_2^{\nu}}
Here $S(x_1,x_2)$ is the length of a geodesic connecting boundary points $x_1$ and $x_2$. By arguments analogous to the ones used in \cite{Czech:2015qta}, one can conjecture the kinematic metric:
\myeq{ds_{\rm{kin}}^2=\frac{\partial^2 S(x_1,x_2)}{\partial x_1^{\mu}\partial x_2^{\nu}}\,dx_1^{\mu}dx_2^{\nu}}
When the bulk is pure AdS, this reduces to eq.~(\ref{eq:higherdmetric}). It would be interesting to use this structure to organize CFT operators and construct duals of bulk Radon transforms in general geometries.  It may be fruitful as well to use intertwinement \eqref{ads3-intertwinement-laplacian} as a principle to determine the metric on kinematic space.

\subsection{Entanglement and Gravity}
The majority of our discussion focused on operators with zero spin, but our framework is more general. OPE blocks for operators with spin correspond to \emph{longitudinal Radon transforms}: integrals over geodesics or minimal surfaces of the bulk tensor field contracted with the tangent vector(s). A special example briefly discussed in the text is the OPE block of the stress tensor family. We showed that this is precisely the vacuum modular Hamiltonian, a central object in the study of entanglement entropy. Its holographic dual is readily provided by the Ryu-Takayanagi relation: it is the perturbation in the area of the minimal surface or, in our language, the longitudinal Radon transform of the metric perturbation. 

We will capitalize on this observation in an upcoming publication. In brief, the intertwinement property of the Laplacian holds for the vector Radon transform exactly as it does for the scalar version. Applied to the modular Hamiltonian, this fact implies the equivalence of linearized Einstein's equations about the AdS vacuum and the kinematic wave equation satisfied by the modular Hamiltonian.

\subsection{De Sitter Holography?}
A curious and noteworthy fact is the appearance of de Sitter space in the study of OPE blocks. For CFTs in $d=2$, the kinematic geometry is $\rm{dS}_2\times \rm{dS}_2$ and the two components decouple entirely. But even in dimensions $d>2$, the kinematic space of minimal surfaces at a fixed boundary time has the geometry of $\rm{dS}_{d+1}$ (an observation made in \cite{PhysRevLett.116.061602}, see also \cite{solanes}). The latter made its appearance in our work as the relevant kinematic space for OPE blocks of conserved currents, since conservation could be combined with the kinematic Klein-Gordon equation to yield a differential equation on the kinematic space of a time-slice. Moreover, the boundary conditions imposed on us by the OPE coincide with the ``future boundary conditions'' \cite{Anninos:2012qw} introduced in the context of the dS/CFT correspondence \cite{dscft} (see Appendix \ref{propagatorchoice}).

It is, therefore, legitimate to raise the question: can OPE blocks propagating in kinematic space serve as the starting point for constructing QFT on de Sitter space? And what is the relation, if any, of the kinematic space formulation for CFTs and the dS/CFT conjecture? The kinematic Klein-Gordon equation derived in the text can be interpreted as an equation of motion for low energy de Sitter fields at leading order in $1/N$. At sub-leading orders, however, the non-trivial CFT $n$-point functions are in tension with the ``free'' propagation implied by the Klein-Gordon equation. In result, ``local'' kinematic fields receive $1/N$ corrections in a way that is essentially identical to the procedure HKLL introduced in the study of AdS holography. These corrections are chosen in order to restore locality in kinematic space---at the expense of introducing local interaction vertices in the equations of motion. Preliminary results indicate that there is no fundamental obstruction to implementing these corrections order by order in $1/N$. It is worth pointing out, however, that correcting the OPE blocks to obtain local dynamics in kinematic space does \emph{not} coincide with the $1/N$ corrections required by AdS locality, which we discussed in Sec.~\ref{dis:interactions}. It will be exciting to explore whether kinematic space hides interesting lessons about de Sitter holography.

\acknowledgments{We thank Ahmed Almheiri, Vijay Balasubramanian, Alexandre Belin, Jan de Boer, Ethan Dyer, Glen Evenbly, Michael Freedman, Micha{\l} Heller, Eliot Hijano, Henry Maxfield, Don Marolf, Rob Myers, Eric Perlmutter, Vladmir Rosenhaus, Edgar Shaghoulian, Steve Shenker, Eva Silverstein, Joan Sim{\'o}n, Leonard Susskind, Gunther Uhlmann, Herman Verlinde and Guifr{\'e} Vidal for useful discussions. BC, LL, SM, and JS thank the KITP, the organizers of the ``It from Qubit" Tensor Network Meeting (supported by the Simons Foundation), and the organizers of the KITP Follow-On Program (supported in part by the National Science Foundation under Grant No. NSF PHY11-25915), as well as the organizers of the ``Quantum Information Theory in Quantum Gravity II" meeting and the Perimeter Institute for Theoretical Physics (supported by the Government of Canada through Industry Canada and by the Province of Ontario through the Ministry of Research and Innovation). LL thanks the organizers of the TASI 2015. SM was supported in part by an award from the Department of Energy (DOE) Office of Science Graduate Fellowship Program. BM is supported by The Netherlands Organisation for Scientific Research (NWO).}

\appendix

\section{Homogeneous Spaces} \label{homogeneous}

Let us understand the various kinematic spaces as homogeneous spaces. This will illuminate why the
conformal Casimir equation of Sec.~\ref{opeblocks} is the Laplace
equation in kinematic space. 

As a warm-up, let us consider $\mathrm{AdS}_{n+1}$ as a homogeneous
space. Its isometry group $\mathrm{SO}\left(n,2\right)$ acts transitively---in other words, there is no distinguished point
in $\mathrm{AdS}$. This implies that $\mathrm{AdS}$ can be written
as a coset space 
\begin{equation}
\mathrm{AdS}_{n+1}=\frac{\mathrm{Isom}\left(\mathrm{AdS}_{n+1}\right)}{\mathrm{Stab}\left(\mathrm{point}\in\mathrm{AdS}\right)}=\frac{\mathrm{SO}\left(n,2\right)}{\mathrm{SO}\left(n,1\right)}\,,
\end{equation}
where $\mathrm{Stab}\left(\mathrm{point}\in\mathrm{AdS}\right)$ denotes the stabilizer subgroup of a point.

Recall that a Lie group $G$ has a distinguished bi-invariant metric given by the Cartan-Killing form.\footnote{In fact, when $G$ is simple, it is the unique such metric. If $G$
is only semi-simple, then there is a free coefficient for each additional
factor. For the cases of our interest, these coefficients are fixed
by discrete symmetries relating the group factors.} With this metric, the quadratic Casimir operator $C_{G}$ is identified
with the Laplacian $\square_{G}$. 

A coset space $G/H$ also inherits a distinguished metric from
$G$. In fact, the Laplacian on $G$ can be written as 
\begin{equation}
\square_{G}=\square_{G/H}+\square_{H}.
\end{equation}
If we now consider a function on $G/H$, it can be lifted to a function
on $G$ that is constant on $H$-orbits. Hence, it is annihilated
by $\square_{H}$. Putting these two facts together, the Casimir operator
$C_{G}$ is represented on functions on $G/H$ by the Laplacian $\square_{G/H}$.
Applying this result to $\mathrm{AdS}_{n+1}$, we find that the Casimir
element $L^{2}$ of the conformal group is identified with the $\mathrm{AdS}$
Laplacian. 
\begin{equation}
L^{2}\underset{\text{AdS scalar}}{\longleftrightarrow} -\square_{\mathrm{AdS}} \label{l2-representation}
\end{equation}
where the minus sign is conventional.  This fact, which was first noted by \cite{Balasubramanian:1998sn}, implies that the quadratic Casimir is given by $C=-m^2$ for an AdS field obeying the wave equation $\left(\square-m^2\right)\phi=0$.  In other words, free AdS fields live in irreducible representations of the global conformal group.

Now let us apply these methods to $\mathcal{K}_{g}$,
the kinematic space of geodesics in a manifold. The kinematic space for $\mathrm{AdS}_{n+1}$
is also homogeneous; any pair of spacelike separated points
can be mapped to any other by a conformal transformation. Let us then
write it as a coset space as before: 
\begin{equation}
\mathcal{K}_{g}\left(\mathrm{AdS}_{n+1}\right)=\frac{\mathrm{Isom}\left(\mathrm{AdS}_{n+1}\right)}{\mathrm{Stab}\left(\mathrm{geodesic}\in\mathrm{AdS}\right)}=\frac{\mathrm{SO}\left(n,2\right)}{\mathrm{SO}\left(n-1,1\right)\times\mathrm{SO}\left(1,1\right)}.
\end{equation}
The stabilizer of a geodesic has two factors, the $\mathrm{SO}\left(n-1,1\right)$
corresponding to boosts about the geodesic and the $\mathrm{SO}\left(1,1\right)$
corresponding to translations along the geodesic. At the boundary,
these correspond to rotations and scalings that fix
the two endpoints. 

Applying the same logic as above, we find:
\begin{equation}
L^{2}\underset{\text{KS scalar}}{\longleftrightarrow}\square_{\mathcal{K}_{g}}\,.
\end{equation}
The relative sign with (\ref{l2-representation}) can be checked, for instance,
by comparing the distance by which a point in AdS and a point in kinematic
space are displaced under a finite time translation.

The surface kinematic space $\mathcal{K}_{s}$ is also a homogeneous
space. It too can be written as a coset space:
\begin{equation}
\mathcal{K}_{g}\left(\mathrm{AdS}_{n+1}\right)=\frac{\mathrm{Isom}\left(\mathrm{AdS}_{n+1}\right)}{\mathrm{Stab}\left({\rm surface}\in\mathrm{AdS}\right)}=\frac{\mathrm{SO}\left(n,2\right)}{\mathrm{SO}\left(n-1,1\right)\times\mathrm{SO}\left(1,1\right)}.
\end{equation}
This again implies: 
\begin{equation}
L^{2}\underset{\text{KS scalar}}{\longleftrightarrow}\square_{\mathcal{K}_{s}}.
\end{equation}

\section{OPE Blocks from Shadow Operators}

\label{app:shadows} 

In this section, we will derive the result (\ref{eq:smeared})
in an additional way, related to the shadow operator formalism
of \cite{Ferrara:1972xe,Ferrara:1973vz,Ferrara20082,Ferrara2008,SimmonsDuffin:2012uy}. 

In (\ref{eq:local-ope}), we expanded the operator product in terms
of quasiprimaries $\mathcal{O}_{k}\left(x\right)$ and
their descendants $\partial_{\mu}\cdots\partial_{\nu}\mathcal{O}_{k}\left(x\right)$ at one point. These operators form a complete basis of CFT operators, which we may call a \emph{local} basis. 

Equivalently, we could consider an alternative basis of operators
consisting of the quasiprimaries $\mathcal{O}_{i}\left(x\right)$
at \emph{every} point, without any descendants. We will call this
the \emph{global} basis. The result (\ref{eq:cft2-smeared-ope}) shows
how to expand the product of $\mathrm{CFT}_{2}$ scalars in this basis
for the case of a scalar contribution. 
The two bases are related by Taylor expansion: 
\begin{eqnarray}
\mathcal{O}_{i}\left(x\right) & = & e^{-ix\cdot P}\mathcal{O}_{i}\left(0\right)e^{ix\cdot P}\nonumber \\
 & = & \left(1+x^{\mu}\partial_{\mu}+\frac{1}{2}x^{\mu}x^{\nu}\partial_{\mu}\partial_{\nu}+\ldots\right)\mathcal{O}_{i}\left(0\right).
\end{eqnarray}

We will now derive the result (\ref{eq:cft2-smeared-ope}) in a different
way that allows us to quickly generalize to higher dimensions and
tensor OPE contributions. First, we make an ansatz for the OPE as
a smeared operator 
\begin{equation}
\mathcal{O}_{i}\left(x\right)\mathcal{O}_{j}\left(y\right)\stackrel{?}{=}\sum_{k}C_{ijk}\int d^{d}z\,F_{ijk}\left(x,y,z\right)\,\mathcal{O}_{k}\left(z\right)\,,
\label{eq:global-ope-ansatz}
\end{equation}
where $F_{ijk}\left(x,y,z\right)$ is a ``smearing function'' that
should also be determined by conformal invariance. 

Let us examine how the function $F_{ijk}$ 
transforms under an arbitrary global conformal transformation. Noting
the transformation property (\ref{eq:local-operator-transformation})
of the local operators and the transformation $d^{d}z\rightarrow\Omega\left(z^{\prime}\right)^{-d}\,d^{d}z^{\prime}$
of the integration measure, we find that $F_{ijk}$ must transform
as:
\begin{equation}
F_{ijk}\left(x,y,z\right)\rightarrow\Omega\left(x^{\prime}\right)^{\Delta_{i}}\Omega\left(y^{\prime}\right)^{\Delta_{j}}\Omega\left(z^{\prime}\right)^{d-\Delta_{k}}F_{ijk}\left(x^{\prime},y^{\prime},z^{\prime}\right).
\end{equation}
This transformation rule is exactly the same as a vacuum three-point
function of three local operators, which is fixed completely by conformal
invariance. Hence, $F_{ijk}$ must be of the form 
\begin{equation}
F_{ijk}\left(x,y,z\right)\propto\left\langle \mathcal{O}_{i}\left(x\right)\mathcal{O}_{j}\left(y\right)\tilde{\mathcal{O}}_{k}\left(z\right)\right\rangle 
\end{equation}
where $\tilde{\mathcal{O}}_{k}$ is a ``fake'' operator of dimension
$\tilde{\Delta}_{k}=d-\Delta_{k}$ which is not necessarily present
in the theory, but used only as a formal device. These \emph{shadow
operators} have been studied previously in \cite{Ferrara:1972xe,Ferrara:1973vz,Ferrara2008,Ferrara20082,SimmonsDuffin:2012uy}. 

Altogether, the OPE can be written as:
\begin{equation}
\mathcal{O}_{i}\left(x\right)\mathcal{O}_{j}\left(y\right)=\sum_{k}C_{ijk}n_{ijk}\int d^{d}z\,\left\langle \mathcal{O}_{i}\left(x\right)\mathcal{O}_{j}\left(y\right)\tilde{\mathcal{O}}_{k}\left(z\right)\right\rangle \,\mathcal{O}_{k}\left(z\right)\,,
\end{equation}
where $n_{ijk}$ is a normalization factor that can be fixed by taking the coincident limit. This gives us a succinct expression
for a general OPE block: 
\begin{equation}
\mathcal{B}_{k}^{ij}\left(x,y\right)=
n_{ijk} \,\left|x-y\right|^{\Delta_{i}+\Delta_{j}}
\int d^{d}z\,\left\langle \mathcal{O}_{i}\left(x\right)\mathcal{O}_{j}\left(y\right)\tilde{\mathcal{O}}_{k}\left(z\right)\right\rangle \,\mathcal{O}_{k}\left(z\right)\label{eq:general-smeared-ope}
\end{equation}
Let us now compare (\ref{eq:general-smeared-ope})
to the expression (\ref{eq:cft2-smeared-ope}), which we
obtained using kinematic space methods. 

Recall that the CFT scalar field 3-point function has the form: 
\begin{equation}
\left\langle \mathcal{O}_{i}\left(x\right)\mathcal{O}_{j}\left(y\right)\mathcal{O}_{k}\left(z\right)\right\rangle =\frac{C_{ijk}}{\left|x-y\right|^{\Delta_{i}+\Delta_{j}-\Delta_{k}}\left|y-z\right|^{\Delta_{j}+\Delta_{k}-\Delta_{i}}\left|x-z\right|^{\Delta_{i}+\Delta_{k}-\Delta_{j}}}.
\end{equation}
If we set $\Delta_{i}=\Delta_{j}$ above and replace $\mathcal{O}_{k}$
with the ``shadow operator'' of dimension $\tilde{\Delta}_{k}=d-\Delta_{k}$,
we indeed find 
\begin{equation}
\left|x-y\right|^{\Delta_{i}+\Delta_{j}}\left\langle \mathcal{O}_{i}\left(x\right)\mathcal{O}_{j}\left(y\right)\tilde{\mathcal{O}}_{k}\left(z\right)\right\rangle \propto\left(\frac{\left|x-z\right|\left|z-y\right|}{\left|x-y\right|}\right)^{\Delta_{k}-d}.
\end{equation}
Thus, the smearing function obtained with the shadow formalism matches the earlier result.   We also learn from comparison with (\ref{eq:cft2-smeared-ope}) that the shadow 3-point function can be thought of as a bulk-to-boundary propagator in kinematic space. In this sense, the shadow operator should be thought of as a source for the kinematic space field.

It is also immediately clear how to generalize to tensor contributions.
We simply add to the shadow operator indices that transform oppositely
to those on $\mathcal{O}_{k}$: 
\begin{equation}
\mathcal{B}_{k}^{ij}\left(x,y\right)=
n_{ijk}\,\left|x-y\right|^{\Delta_{i}+\Delta_{j}}
\int d^{d}z\,\left\langle \mathcal{O}_{i}\left(x\right)\mathcal{O}_{j}\left(y\right)\tilde{\mathcal{O}}_{k\,\mu\nu\ldots}\left(z\right)\right\rangle \,\mathcal{O}_{k}^{\mu\nu\ldots}\left(z\right).\label{most-general-smeared-representation}
\end{equation}
Note that the integration region is a subtle point, as discussed in Sec.~\ref{higherd}.

\section{Propagator on \texorpdfstring{$\rm{dS}_2 \times \rm{dS}_2$}{dS2 x dS2} and Conformal Block Analytics}
\label{propagatorchoice}

Conformal blocks are computed by a correlation function of OPE blocks (\ref{eq:CBasKSP}), which in kinematic space admits the interpretation of a two-point function or a propagator. Kinematic space, however, is a space of mixed signature, a fact that permits the construction of various inequivalent propagators. It is instructive to understand which choice of propagator makes contact with conformal blocks. For simplicity we restrict our discussion to the two-dimensional case, but analogous statements should hold in any dimension. 

Since the CFT$_2$ kinematic space factorizes, it suffices to understand the propagator $G_{K}$ in one of the de Sitter components, e.g. the one corresponding to the right-moving null coordinates of the bi-locals: $(z_1, z_2)$. The two-point function in $\rm{dS}_2$ is a solution to the differential equation:
\myeq{\left( Q^2-1\right)\frac{d^2}{dQ^2}G_{K}(Q) +2Q\frac{d}{dQ}G_{K}(Q)+m^2G_{K}(Q)=0}
where:
\myeq{Q=\frac{\ell_A^2+\ell_B^2-(z_A-z_B)^2}{2\ell_A\ell_B}} 
is the de Sitter invariant distance between the kinematic points $A$ and $B$ and the coordinates used are the co-moving coordinates:
\myal{l_A= \frac{z_2^{(A)}-z_1^{(A)}}{2} \\
z_A=\frac{z_2^{(A)}+z_1^{(A)}}{2}}
and similarly for point $B$.

In order to specify a unique solution to this equation, we have to introduce an extra condition for the propagator. This condition comes from imposing the appropriate boundary conditions. Recalling the boundary conditions for the OPE blocks discussed in Sec.~\ref{opeblocks}, we require that when any of the kinematic points (e.g. point $A$) approaches the asymptotic boundary $\ell_A\rightarrow 0$, the propagator should fall off as
\myeq{G_{K}\sim \ell_A^h,}
where $h=\frac{1}{2}+\sqrt{\frac{1}{4}-m^2}$.
These boundary conditions are known in the dS/CFT literature as ``future boundary conditions'' \cite{Anninos2012} and there exists a unique solution that obeys them:
\myeq{G_{K}(Q)= \left( \frac{2}{1-Q}\right)^h \, _2F_1\left(h,h,2h,\frac{2}{1-Q}\right) }
This result can be directly matched with the explicit form of global conformal blocks in two dimensions, which reads: \myeq{g_{k|1234}= z^{h_k} \,_2F_1 \left(h_k,h_k,2h_k, z\right). \label{eq:block2D} }
In this expression, $z$ is the cross ratio constructed from the right-moving null coordinates of the four points: 
\myeq{z= \frac{z_{21}z_{43}}{z_{31}z_{42}}.}

It is worth noting that this is \emph{not} a causal propagator in de Sitter space. It does not compute a scalar two-point function in any of the normalizable de Sitter $\alpha$--vacua. This can be readily understood by noticing that this propagator does not reproduce the expected flat space singularity structure at small distances: Besides the expected singularities at null separated points, (\ref{eq:block2D}) also has antipodal singularities. This property is, however, consistent with the elliptic character of the kinematic de Sitter space \cite{schrodinger, ellipticds2}.

\begin{figure}
\centering{\includegraphics[width=0.9\textwidth]{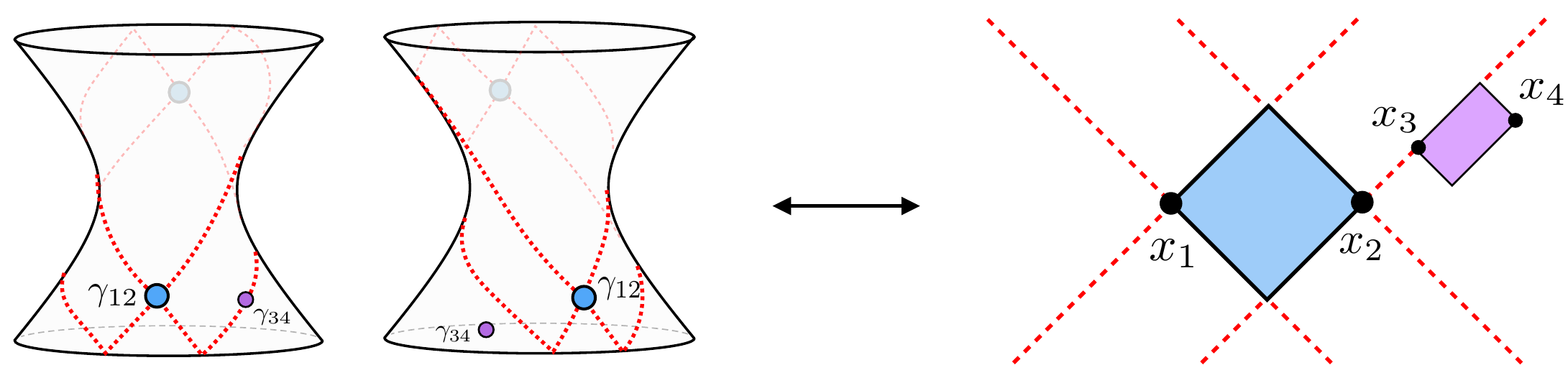}
\caption{The conformal block has singularities whenever two points are lightlike-separated from each other.  This gives rise to a rich singularity structure of the kinematic space propagator, which computes the conformal block. \label{fig:conformal-block-singularities}}}
\end{figure}

\paragraph{Discrete symmetries of CFT$_2$ kinematic space} The full four-dimensional kinematic space, defined as the space of pairs of points $q=(x^{\mu}_1,x^{\mu}_2)$, is symmetric under the discrete transformation $P$ that exchanges the two:
\myeq{q \rightarrow P q \,\,\,\Rightarrow\,\,\, (x^{\mu}_1,x^{\mu}_2)\rightarrow (x^{\mu}_2,x^{\mu}_1)}
Using null CFT coordinates to parametrize the boundary points $x_i^{\mu}=(z_i,\bar{z}_i)$, the exchange transformation $P$ can be expressed as the product of two operators $\mathcal{P}$ and $\bar{\mathcal{P}}$ which exchange the right-moving  and left-moving null coordinates of the bi-local, respectively:
\myeq{P=\mathcal{P}\bar{\mathcal{P}}} 

A more subtle point is that the kinematic space of CFT$_2$ is symmetric under the individual right-moving and left-moving exchange maps, $\mathcal{P} \text{ and } \bar{\mathcal{P}}$. To see this, first recall that for a 2-dimensional boundary, kinematic space is the space of causal diamonds. Exchanging the endpoints of a diamond via an application of $P$, obviously leaves it invariant. However, there exist another discrete transformation that leaves the diamond unaffected: mapping the two space-like separated points that define it to the top and bottom tips of the diamond. This is precisely the effect of the individual $\mathcal{P}$ and $\bar{\mathcal{P}}$ (see Fig. \ref{fig:conformal-block-singularities}). In the geometric picture, individual $\mathcal{P}$ and $\bar{\mathcal{P}}$ transformations correspond to antipodal maps on the two independent $\rm{dS}_2$ components of kinematic space. 

\paragraph{Singularities}
Conformal blocks become singular when two boundary insertions become null separated in the CFT spacetime. More specifically, the right-moving conformal block, which we are interested in, captures the singularities that come from right-moving null alignments of the 4 points:
\myal{ z_1&=z_3 \,\,\text{or}\,\, z_2=z_4 \,\,\,\Rightarrow \,\,\, z=\infty\\
z_1&=z_4 \,\,\text{or}\,\, z_2=z_3 \,\,\,\Rightarrow \,\,\, z=1}

These singularities from the perspective of kinematic space correspond to invariant distances $Q=1$ and $Q=-1$, respectively. The singularity at $Q=1$ appears when the kinematic points defined by the corresponding bilocals are null separated according to the kinematic causality explained in detail in Section \ref{ads3kin}. This is expected since $\rm{dS}_2$ is a Lorentzian manifold and therefore propagators have null singularities. 

The $Q=-1$ singularity occurs when $z_2=z_3$ or $z_1=z_4$, which means that the kinematic points define entirely disconnected causal patches. Recall, however, that under a right-moving antipodal map $\mathcal{P}$ the bilocal $(z_3,z_4)$ becomes $(z_4,z_3)$. Therefore, when $Q=-1$ the bilocal $(z_1,z_2)$ is null separated from the antipodal point $\mathcal{P}(z_3,z_4)$. Since the antipodal point corresponds to the same causal diamond in the CFT, the existence of the $Q=-1$ singularity is dictated by the discrete symmetry of kinematic space.

\bibliographystyle{jhep}
\bibliography{bibliography}

\end{document}